\pdfoutput=1
\documentclass[twocolumn, twocolappendix]{aastex631}

\usepackage{wrapfig}
\usepackage{amsmath}
\usepackage{tikz}
\newcommand{\tikzcircle}[2][red,fill=red]{\tikz[baseline=-0.5ex]\draw[#1,radius=#2] (0,0) circle ;}%

\begin{document}

\title{Resolving Pleiades binary stars with Gaia and speckle interferometric observations}

\author[0000-0001-5038-0089]{Dmitry Chulkov}
\affiliation{Institute of Astronomy of the Russian Academy of Sciences (INASAN) \\
119017, Pyatnitskaya st., 48, Moscow, Russia ; \href{mailto:chulkovd@gmail.com}{chulkovd@gmail.com}}

\author[0000-0003-0647-6133]{Ivan Strakhov}
\affiliation{Sternberg Astronomical Institute, Lomonosov Moscow State University\\ 119992 Universitetskii prospekt 13, Moscow, Russia}

\author[0000-0003-1713-3208]{Boris Safonov}
\affiliation{Sternberg Astronomical Institute, Lomonosov Moscow State University\\ 119992 Universitetskii prospekt 13, Moscow, Russia}



\begin{abstract}

The Pleiades is the most prominent open star cluster visible from Earth and an important benchmark for simple stellar populations, unified by common origin, age, and distance. Binary stars are its essential ingredient, yet their contribution remains uncertain due to heavy observational biases. A resolved multiplicity survey was conducted for a magnitude-limited $G<15^{\rm mag}$ sample of 423 potential cluster members, including sources with poorly fitted astrometric solutions in Gaia DR3. Speckle interferometric observations at the 2.5 meter telescope of SAI MSU observatory were combined with Gaia data, enabling the identification of 61 resolved binary or multiple systems within the 0.04 -- 10 arcsec (5 -- 1350 au) separation range. With speckle observations, we discovered 21 components in 20 systems. The existence of a Merope (23 Tau) companion is confirmed after several previous unsuccessful attempts. We show that the Gaia multipeak fraction is a strong predictor of subarcsecond multiplicity, as all sources with  {\it ipd\_frac\_multi\_peak}~$>4$\% are successfully resolved.  We found that 10\% of Pleiades stars have a companion with a mass ratio $q>0.5$ within projected separation  of $27< s <1350$ au, and confirm a deficit of wide binaries with $s>300$~au.  \textcolor{black}{An observed dearth of wide pairs with large mass ratio ($q>0.55$) may imprint the transition from hard to soft binaries regime at the early stages of cluster evolution.} The total binary fraction for $q>0.5$ systems is extrapolated to be around 25\%. 
\end{abstract}



\section{Introduction} \label{sec:intro}

The world of binary and multiple stars is incredibly diverse and is of invaluable importance for our understanding of the Universe. Despite major efforts in recent years, the statistical properties of binary populations and their variations in different environments remain controversial, complicated by numerous selection effects in observational data \citep{2024PrPNP.13404083C}. The exact mechanisms of multiple star formation, along with their relative contributions, remain under discussion. The agreement between different models and observed statistics serves as a natural test of their validity  \citep{2023ASPC..534..275O}. The topic of stellar multiplicity is intimately related to the study of open clusters, as most stars are ultimately born in dense stellar environments, and their dissolution gives rise to one of the binary formation mechanisms \citep{2010MNRAS.404.1835K}. As coeval and chemically homogeneous groups, open clusters serve as benchmarks for Galactic structure and are useful for developing stellar evolution models \citep{2024NewAR..9901696C}. Comparing observed multiplicity statistics in different clusters can reveal dependencies on external factors, although this task is severely complicated by observational biases. Recently, the Gaia mission \citep{2016A&A...595A...1G} has opened a new chapter for multiplicity studies, and more significant results are anticipated with upcoming data releases \citep{2024NewAR..9801694E}. At present, combining Gaia data with ground-based high-resolution observations increases the number of detected systems and enables less biased multiplicity statistics \citep{2023AJ....165..180T}.

The Pleiades is one of the most prominent and arguably beautiful naked-eye objects in the night sky, known since prehistoric times \citep{2001EM&P...85..391R, 2008arXiv0810.1592S,2021acas.book..223N}. Being the open cluster richest in number of members within 400 pc from the Sun \citep{2023MNRAS.526.4107P}, it has been an important benchmark. Numerous studies have examined multiplicity for its members with dedicated adaptive optics or spectroscopic observations \citep{1997A&A...323..139B, 2018AJ....155...51H, 2021ApJ...921..117T} or through analysis of photometric data \citep{2022AJ....163..113M,2023MNRAS.525.2315A}. This paper focuses on the analysis of the resolved binary population based on Gaia DR3 main source catalogue \citep{2021A&A...649A...1G} and dedicated speckle interferometric observations (Section~\ref{observations}). 
If trends for solar-type stars \citep{2010ApJS..190....1R} and M dwarfs \citep{2019AJ....157..216W} in the solar neighborhood are relevant for the Pleiades, the peak of companion frequency falls into 0.1 -- 1 arcsec separation range, suitable for high angular resolution observations.

In Section \ref{sample}, we describe the observational sample and introduce the isochrone used for mass and mass ratio estimation in Section \ref{Mass}. The complexity of resolved and unresolved Gaia sources is discussed in Section \ref{resolved_types}, with further analysis of Gaia-resolved systems in Section \ref{Gaia-resolved}. Speckle observations and data reduction are described in Section \ref{observations}, and the discoveries of previously unreported companions are presented in Section \ref{results}. Binary fraction and statistical distributions are analyzed in 
Section \ref{statistics}. 
The key results are summarized in Section~\ref{summary}. The Appendix includes information on insecure cluster members, multiple stars, a complete log of observations for resolved binaries, \textcolor{black}{and obtained detection limits for all entries (Sections \ref{Membership}, \ref{Multiples}, and \ref{log})}. Throughout the paper, we use the Washington Double Star Catalog  \citep{2001AJ....122.3466M} for system designations, introducing new identifiers based on J2000 coordinates when necessary.

\section{Pleiades sample \& isochrone} \label{sample}

\begin{figure}{}
\includegraphics[width=0.47\textwidth]{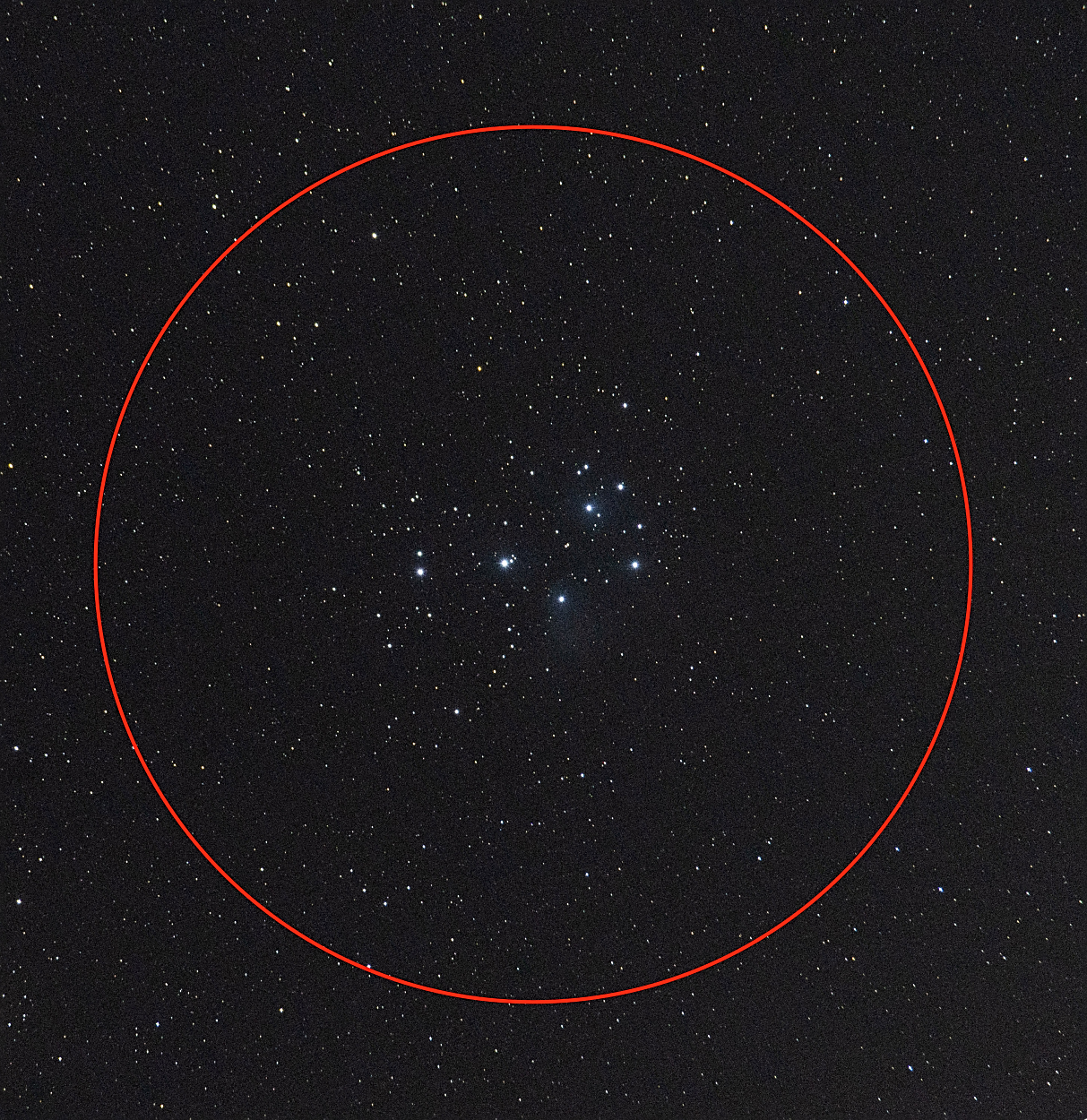}
\caption{The Pleiades (Messier 45, Melotte 22 \citep{1781cote.rept..227M, 1915MmRAS..60..175M}). A red circle encompasses the surveyed area with a radius of 2 degrees. Photo by Dmitry Trushin.}
\label{fig:photo} 
\end{figure}

\textcolor{black}{The entire list of sample objects is presented in table 2 from \cite{2024AJ....168..156C} and is directly accessible through the \href{https://vizier.cds.unistra.fr/viz-bin/VizieR?-source=J/AJ/168/156}{VizieR} service}. Gaia DR3 \citep{2021A&A...649A...1G} sources with $G<15^{\rm mag}$ within the two degree radius shown in Figure \ref{fig:photo} are considered. Cluster members are selected based on proper motion, parallax, and radial velocity from various datasets. 
A special effort is taken to include the often neglected sources with poorly-fitted astrometric solutions in Gaia DR3, often indicative of non-single stars. Overall, the sample comprises 409 probable members and 14 sources with disputed membership. A $G=12^{\rm mag}$ threshold splits the sample evenly into bright (204 entries) and faint subsamples. 

\begin{figure}[t] 
\includegraphics[width=.5\textwidth]{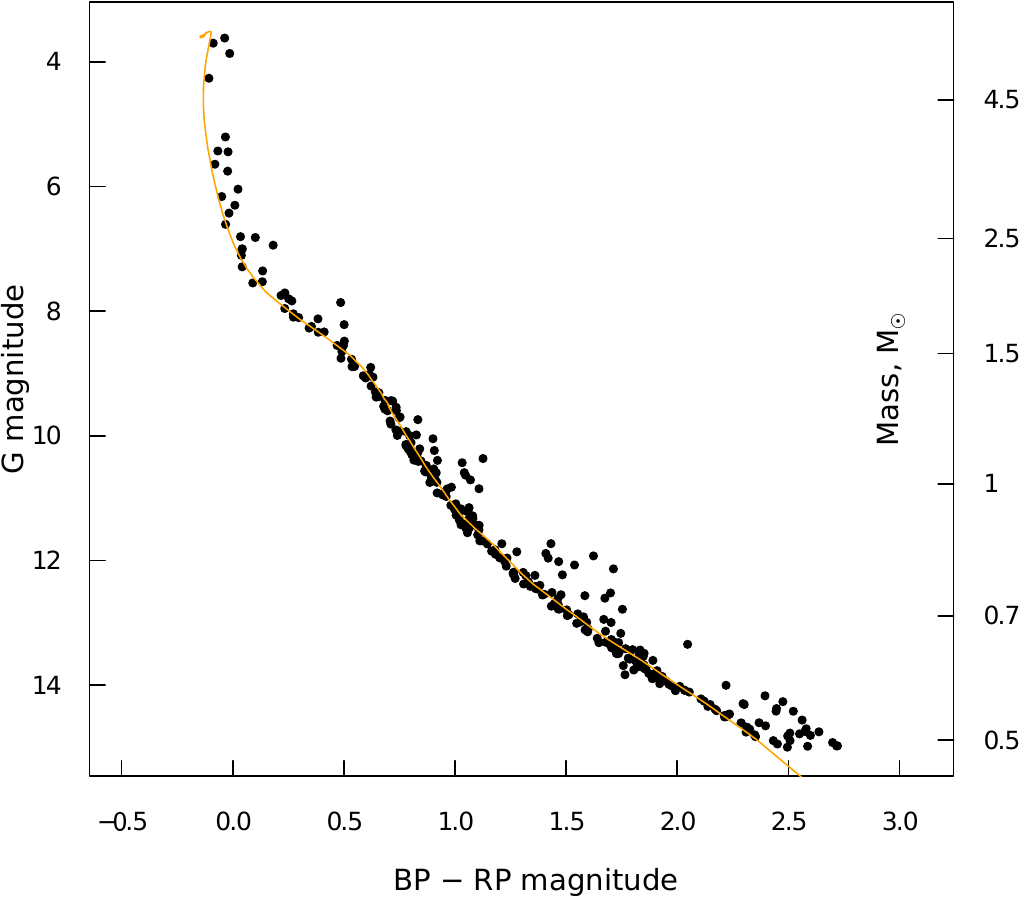}
\caption{PARSEC (version 2.0;  \cite{2012MNRAS.427..127B, 2022A&A...665A.126N}) isochrone with rotation $\Omega/\Omega_c=0.3$ (inclination $60^{\circ}$) for 100 Myr age and [Fe/H] = 0.05, plotted against sample Pleiades stars.  A distance of 135 pc ($5.65^{\rm mag}$ modulus) with an extinction of $A_V=0.15^{\rm mag}$ is applied. The mass of a single star is shown along the right axis.} 
\label{fig:PARSEC} 
\end{figure}

Measuring mass for an isolated star is challenging \citep{2021A&ARv..29....4S}, but for open cluster members, it depends directly on apparent magnitude. Most sample objects fall into an isochrone blindspot region, where the age dependence is minimal \citep{2024A&A...690A..16R}. The 100 Myr PARSEC isochrone \citep{2012MNRAS.427..127B, 2022A&A...665A.126N} reasonably agrees with the observed sequence in the color-magnitude diagram (Figure~\ref{fig:PARSEC}) within the covered parameter space and is used for mass estimation. Masses for $G>15^{\rm mag}$ ($m\lesssim 0.5M_\odot$) stars are less reliable \citep{2024arXiv241112987W}. To account for the width of the main sequence, we allow a systematic uncertainty of $0.15^{\rm mag}$ to the reported $G$ magnitudes.

\section{Mass estimation}
\label{Mass}

When binary system components appear in Gaia DR3 as separate entries, their masses and mass ratio (${q=m_2/m_1}$, $0<q\le1$) are estimated from $G_1$ and $G_2$ magnitudes. Instead, a close unresolved binary behaves as a single source with the combined flux of both components. For example, a lone solar mass star is expected to have $G=10.77^{\rm mag}$, while an unresolved twin binary of $1.75\ M_\odot$ total mass with identical components would have the same reported  magnitude. The distinction between resolved and unresolved cases, discussed in Section~\ref{resolved_types}, has a limited impact on $q$ estimation.
Magnitudes for primary and secondary stars and combined system's flux are calculated according to the isochrone across a dense grid of masses and mass ratios. The values of $m_1$ and $q$ that provide agreement with the observed $G$ magnitude and observed flux contrast are selected to constrain the parameters of the binary system. 

The median reported $G$ error for sample objects is $0.003^{\rm mag}$. When Gaia magnitudes are known for both components, a $0.01^{\rm mag}$ error is adopted for the contrast $\Delta G = G_2 - G_1$, along with a systematic $0.15^{\rm mag}$ uncertainty for $G_1$. We perform $10^4$ Monte Carlo simulations with normally distributed uncertainties; the 0.15 and 0.85 quantiles define the error margins. The uncertainties of speckle observations are discussed in Section~\ref{errors}.

Speckle observations are conducted with several filters (Section \ref{observations}). Stars with $G<9.5^{\rm mag}$ are normally observed in 50 and 70 nm half-width passbands centered at 550 and 880 nm, respectively. Their response curves are outside the standard photometric systems, so isochrone magnitudes are not provided for them. Fortunately, the center line of the 550 nm filter is close to the $y$ passband \citep{1966ARA&A...4..433S}, while the 880 nm filter closely resembles the UKIDDS $Z$ passband \citep{2006MNRAS.367..454H}, and we use them for calculations. Hereinafter, we refer to the 550 and 880 nm bands as $\hat{y}$ and~$\hat{z}$.  For nine binaries with repeated multicolor observations, the consistency of inferred $q$ with different filters matches the scatter within the same passband, with a pooled standard deviation $\sigma_q\sim0.02$. Moderate photometric accuracy of speckle observations obscures overall precision (Section~\ref{errors}).

\begin{figure}{}
\includegraphics[width=.47\textwidth]{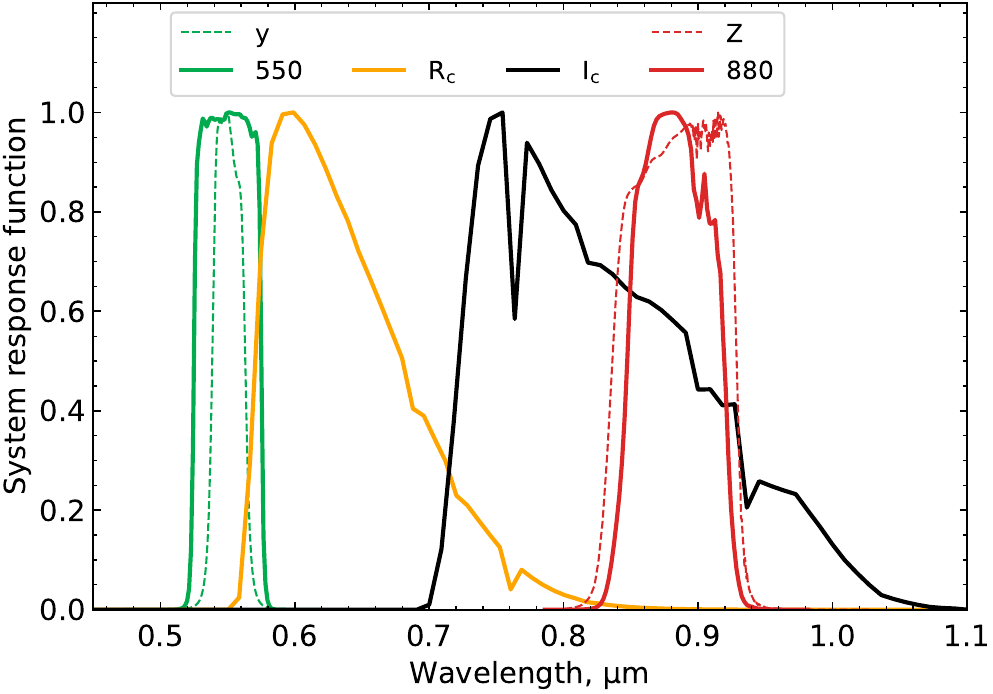}
\caption{System response functions for speckle observations filters, the input of atmospheric airmass at the zenith is taken into account. Solid \textcolor{green}{\bf green}: 550 nm mid-band, 50 nm half-width, designated $\hat{y}$. Solid \textcolor{orange}{\bf orange}: Johnson-Cousins  $R_c$, solid \textbf{black}: Johnson-Cousins  $I_c$. Solid \textcolor{red}{\bf red}: 880 nm mid-band, 70 nm half-width, designated $\hat{z}$. Since the 550 and 880 nm filters are not included in the library used by PARSEC isochrones, Str\"omgren y (\textcolor{green}{dotted green}) and UKIDSS Z (\textcolor{red}{dotted red}) passbands are used as their substitutes.} 
\label{fig:filters} 
\end{figure}

The reliability of isochrone-based mass estimation is validated on the resolved spectroscopic binary WDS~03491+2347, with a mass ratio estimate from the orbital solution by \cite{2020ApJ...898....2T}. For a single Pleiades star, ${G=10.43\pm 0.15^{\rm mag}}$ corresponds to a mass of $1.07 \pm 0.03\ M_\odot$. However, adaptive optics of the Keck Observatory \citep{2013PASP..125..798W} resolved it as a binary with $\rho \sim 0.055$ arcsec separation and ${\Delta J = 0.36 \pm 0.09^{\rm mag}}$ contrast. Such a close pair is unresolved by Gaia, and we search the isochrone grid for a system with a matching combined magnitude and $\Delta J$. A binary with $m_1=0.98 \pm 0.03$ and ${m_2=0.85 \pm 0.04\ M_\odot}$ meets these criteria. The masses implied from the orbital solution alone, without photometric priors, are within the error margin: ${m_1=1.01 \pm 0.04}$ and $m_2=0.87\pm 0.04\ M_\odot$. The $q$ values derived from magnitude contrast in the $H$ and $K$ passbands are also in good agreement (Table~\ref{tab:v1282}). 

\begin{table}
    \centering
    \hspace{-1.3cm}\begin{tabular}{|c|c|c|}
        \hline
        Passband &Contrast, mag  &Mass ratio, $q$ \\ \hline
        $\Delta I$ & $0.71 \pm 0.06$ & $0.862 \pm 0.011$\\ \hline
         $\Delta J$& $0.36 \pm 0.09$ & $0.911 \pm 0.022$ \\
\hline
         $\Delta H$& $0.34\pm0.05$ & $0.895 \pm 0.016$ \\
\hline
         $\Delta K$& $0.32 \pm 0.02$ & $0.897 \pm 0.010$ \\
         \hline
         
         \multicolumn{2}{|c|}{Orbital solution I}&$0.86\pm0.06$\\
                  \hline
                  \multicolumn{2}{|c|}{Orbital solution II}&$0.917\pm0.004$\\
         \hline
    \end{tabular}
    \caption{The derived $q$ for binary system WDS 03491+2347 based on the magnitude contrast in different passbands. $I_c$-band photometry comes from our speckle observations (Section \ref{errors}), $JHK$ photometric data (2019 measurements are used) and orbital solution are from \cite{2020ApJ...898....2T}. Orbital solution II incorporates $JHK$ photometry as a prior. The reported Gaia DR3 magnitude of $G=10.43^{\rm mag}$ corresponds to the combined flux from both components.}
    \label{tab:v1282}
\end{table}

A fraction of Pleiades stars forms triple and higher-order systems. The presence of the third star disturbs the mass ratio estimation. While the treatment of resolved  systems is straightforward, as magnitude difference of components is directly measured, a custom approach is needed for unresolved sources (Section \ref{Multiples}, Appendix). The undetected unresolved companions remain possible in systems currently presumed to be binary.

\section{Resolved and unresolved sources}
\label{resolved_types}

Gaia photometry corresponds to either an individual star or the combined flux from a binary system, depending on circumstances. The $G$ magnitude is derived from a line spread function fit in astrometric field optimized for point sources \citep{2021A&A...649A..11R}, while $3.5\times2.1$ arcsec aperture photometry measures $B_p$ and $R_p$ fluxes \citep{2021A&A...649A...3R}. Pairs with $\rho \lesssim 1^{\prime\prime}$ are affected by blending during scans regardless of orientation, making the reported $B_p$ and $R_p$ magnitudes reflect total system flux rather than individual components. The spatial resolution in $G$ band is better, with a median FWHM in the along-scan direction of about $0.1^{\prime\prime}$ \citep{2016A&A...595A...3F}, creating an instrumental artifact that overestimates $B_p$ and $R_P$ fluxes relative to $G$ \citep{2023A&A...670A..19G}. Whether Gaia resolves a binary depends on the scan direction, with the parameter {\it ipd\_frac\_multi\_peak} ($f_{\rm MP}$) indicating the fraction of observations where more than one peak is detected during the image parameter determination used for $G$ estimation \citep{2021A&A...649A...2L}. 
Gaia multipeak analysis has revealed extragalactic sources with subarcsecond separation \citep{2022NatAs...6.1185M}, and   \cite{2023AJ....165..180T} highlights $f_{\rm MP}$ as a powerful indicator of speckle-resolved multiplicity. 

\begin{figure*}[t!] 
\begin{minipage}{\linewidth}
\includegraphics[width=\linewidth]{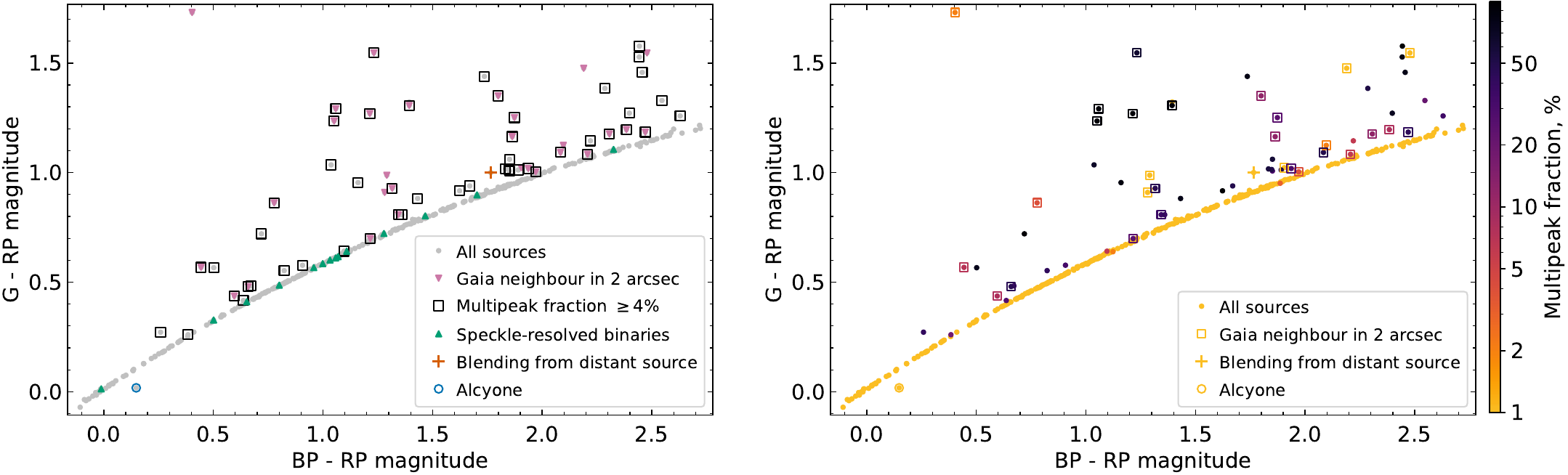}
\end{minipage}
\caption{Color-color diagram for Pleiades sample stars. As discussed in \cite{2024AJ....168..156C}, single stars and close unresolved systems belong to main sequence, while pairs with $\rho \gtrsim 0.1^{\prime\prime}$ appear as outliers due to blending. {\bf Left} -- \textcolor{violet}{\bf Violet $\blacktriangledown$}: another Gaia DR3 source with $G<19^{\rm mag}$ within 2 arcsec is present; black $\square$: source with {\it ipd\_frac\_multi\_peak}$\geq 4\%$; \textcolor{orange}{\bf orange $+$}: bright star (${G=5.75^{\rm mag}}$) at $\rho=8.8^{\prime\prime}$ causes blending; \textcolor{cyan} {\bf cyan $\bigcirc$}: Alcyone; \textcolor{green}{\bf green $\blacktriangle$}: speckle-resolved binary with {\it ipd\_frac\_multi\_peak}~${\le1}$\%. {\bf Right} -- value of {\it ipd\_frac\_multi\_peak} is color-coded, sources with another Gaia DR3 source with ${G<19^{\rm mag}}$ within 2~arcsec are marked with $\square$. All outlying stars either have another Gaia source in close vicinity or {\it ipd\_frac\_multi\_peak} excess.} 
\label{fig:two-color} 
\end{figure*}

Out of 423 sample entries, 54 have ${f_{\rm MP}\geq 4\%}$, all of which belong to resolved systems with $0.07 < \rho < 1.9$ arcsec from speckle observations (Section \ref{observations}). Additionally, they deviate from the main sequence in the color-color diagram (Figure \ref{fig:two-color}) due to blending in $B_p$ and $R_p$ passbands, except for binaries with secondary flux contribution below 1\%. Among 21 sources with $1 \leq f_{\rm MP} \leq3\%$, only three are speckle- or Gaia-resolved with $\rho<2^{\prime\prime}$. However, at least 12 sources with $f_{\rm MP}=0$ are resolved pairs with $\rho<0.3^{\prime\prime}$. These sources lie on the main sequence in the color-color plot, even with significant secondary flux. Conversely, in the color-magnitude diagram, they appear above the main sequence, unless their secondary flux is low. This behavior matches close spectroscopic binaries (figure 7 in \cite{2024AJ....168..156C}), and we adopt that sources with $f_{\rm MP}\leq 1\%$ within $\rho<0.3^{\prime\prime}$ systems are unresolved (blended) in $G$ band. In practice, the transition between resolved and unresolved modes is likely gradual (figure 2 in \cite{2025AJ....169...29S}).  

\section{Gaia-resolved pairs} \label{Gaia-resolved}
\begin{figure*}[h] 
\begin{minipage}{\linewidth}
\includegraphics[width=\linewidth]{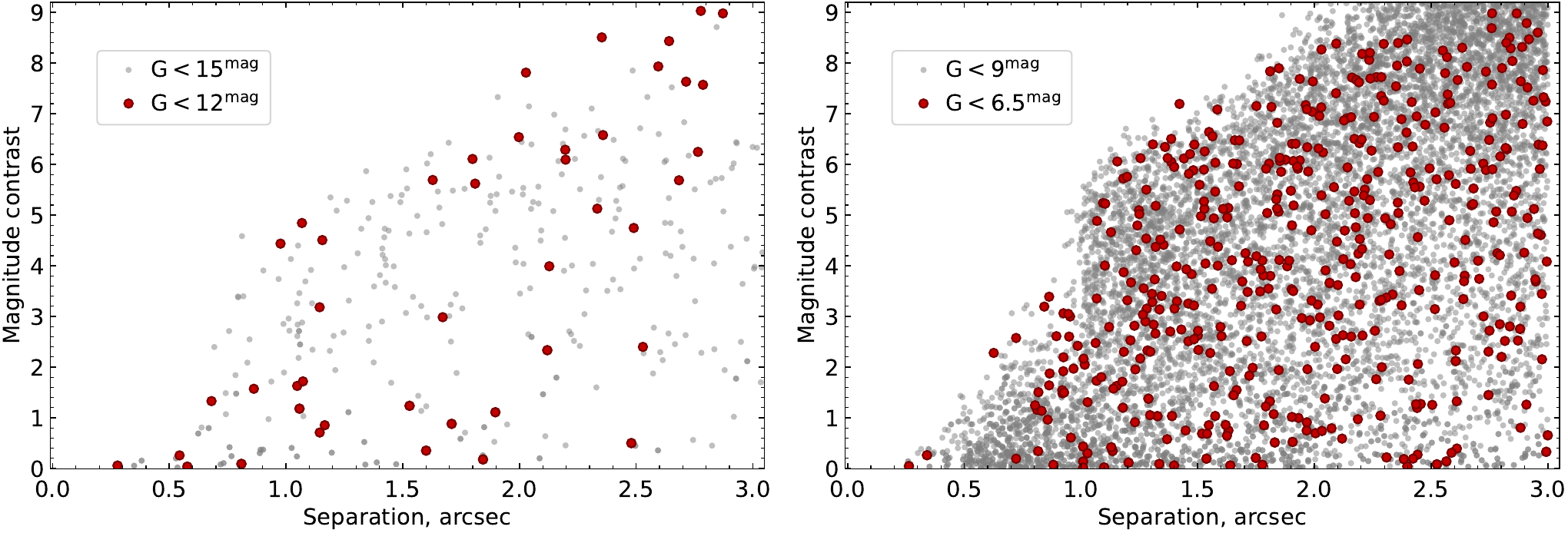} 
\end{minipage}
\caption{Angular separation and magnitude difference $\Delta G= |G_2-G_1|$ for neighboring Gaia DR3 sources. Some pairs are optical. \textbf{Left:} Two degree Pleiades field (cluster membership not required), at least one source has $G<15^{\rm mag}$. Pairs with $G<12^{\rm mag}$ stars are marked with \tikzcircle[black, fill=red]{2pt}. \textbf{Right:} All-sky sample of $G<9^{\rm mag}$ sources, where $G<6.5^{\rm mag}$ stars are marked by \tikzcircle[black, fill=red]{2pt}.}

\label{fig:Gaia} 
\end{figure*}

We start the census of multiplicity with resolved systems where both components appear in Gaia DR3 as separate entries. The minimum separation between sources with $G<15^{\rm mag}$ in the two degree surveyed area, regardless of cluster membership, is 0.28 arcsec (WDS~03494+2456). However, at small separations, only components of similar brightness are resolved, and the detectable contrast is within 1 magnitude until $\rho<0.65^{\prime\prime}$ (Figure \ref{fig:Gaia}). Then the sensitivity rapidly improves to $\Delta G \sim 4^{\rm mag}$ at $\rho \sim 1.0$ arcsec, ensuring at least $q>0.5$ completeness for the considered Pleiades objects. The maximum $\rho$ for a binary star detected by our observational technique that Gaia DR3 fails to resolve is 0.89 arcsec (WDS 03419+2327; Section \ref{observations}). Thus, for $\rho \gtrsim 1^{\prime\prime}$, Gaia is undoubtedly superior to our observations. 

\subsection{Sources with full astrometric solutions}
\label{full_solution}
\begin{figure*}[h] 
\begin{minipage}{\linewidth}
\includegraphics[width=\linewidth]{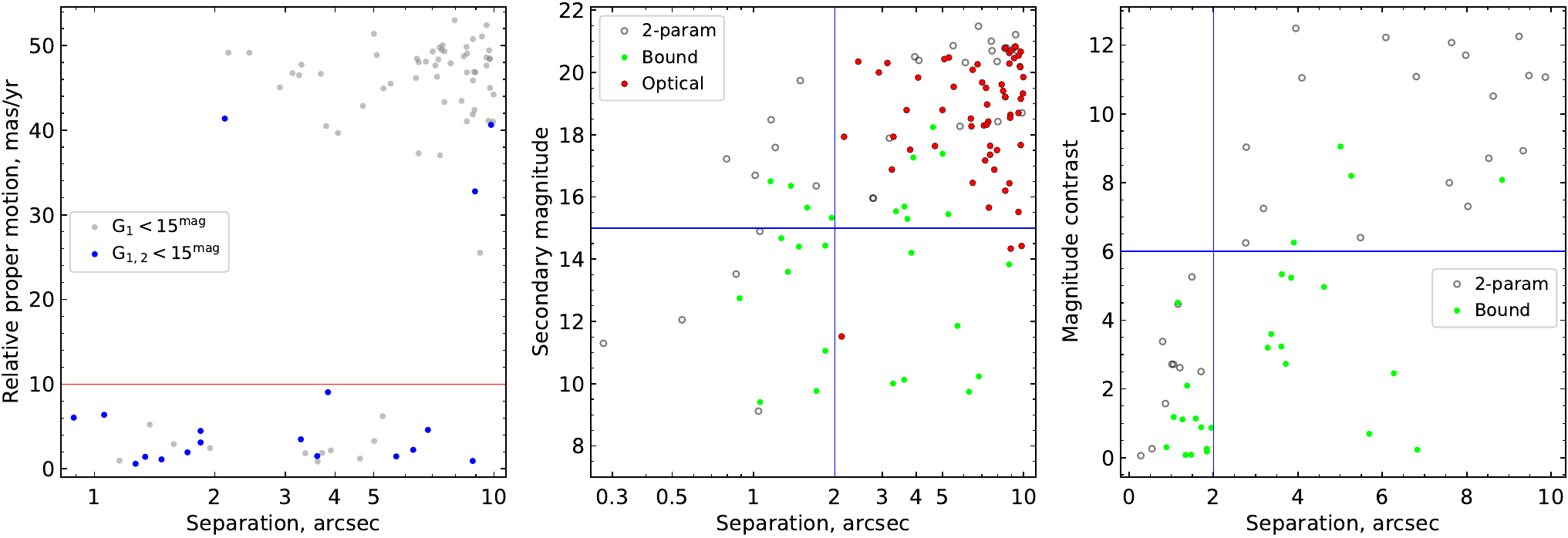}
\end{minipage}
\caption{{\bf Left:} Angular separation and relative proper motion (Equation \ref{eq:pm}) for pairs of Gaia DR3 sources with five- or six-parameter astrometric solutions and $\rho<10$ arcsec. At least one source must have $G<15^{\rm mag}$. Cases where both sources meet this criterion are colored  \textcolor{blue}{\textbf{blue}}. Bound binaries are well separated from optical projections. As expected, most of optical pairs include a faint source. Pairs with $\Delta \mu< 10$ mas~yr$^{-1}$ are considered  \textcolor{green}{\bf bound}, while the rest are \textcolor{red}{\bf optical}. {\bf Middle:} \textcolor{red}{\bf Optical pairs}, \textcolor{green}{\bf bound binary} systems and pairs with unknown proper motion depending on angular separation and magnitude of fainter component. Two-parameter solutions appear only for bright stars at small $\rho$, where the probability of mere projection is low. \textbf{Right:} Magnitude contrast $\Delta G= |G_2-G_1|$ as a function of angular separation. Pairs with full astrometric solutions (Section \ref{full_solution}) are shown with \textcolor{green}{\bf green} filled circles, empty circles represent two-parameter solutions (Section \ref{two-parameter}). Confirmed optical pairs were excluded. For wide binaries with $\rho>2^{\prime\prime}$, the full solutions are available up to $\Delta G<6^{\rm mag}$.}
\label{fig:resolved} 
\end{figure*}

A double star observed in the celestial sphere can be a mere projection, a likelihood that increases with larger separations and fainter components. We restrict our sample to $\rho<10^{\prime\prime}$ pairs (1 arcsec approximates 135 au projected separation for the Pleiades). If both sources have a full astrometric solution in Gaia, the genuine binaries can be distinguished from the optical pairs. The relative proper motion of stars is calculated as follows:

\begin{equation} 
\begin{split}
   &\delta\mu = \sqrt{\Delta \mu^2_{\alpha}+\Delta \mu^2_{\delta}} \\
   &\Delta \mu_\alpha = \cos \delta(\mu_\alpha^A-\mu_\alpha^B)  \\
   &\Delta \mu_\delta = \mu_\delta^A-\mu_\delta^B  
   \label{eq:pm}
\end{split}
\end{equation} 

The principal cause for $\delta \mu$ excess is the orbital motion of stars around their common center of mass. While velocities are higher in close systems, such sources often lack astrometric data. The lowest separation among systems with $\mu$ reported for both components is ${\rho=0.89^{\prime\prime}}$ for WDS 03466+2421. 
The expected orbital period (${\sim\sqrt{\rho^3/m\varpi^3}}$) is on the order of $10^3$ yr, while the circular velocity ($\sim \sqrt{m \varpi / \rho}$) reaches 4.3 km~s$^{-1}$ (6.7~mas~yr$^{-1}$), as for WDS 03463+2411.  This is much lower than the cluster's average $|\mu^c| \sim 50$ mas~yr$^{-1}$, making the relative proper motion of optical pairs larger than for bound binaries.  The observed $\delta \mu$ distribution splits entries into two groups (Figure \ref{fig:resolved}), and we adopt $\delta \mu>10$ mas~yr$^{-1}$ as a threshold for optical projections. The reported $\delta \mu$ can exceed the estimated circular velocity for a given system due to limitations of the single-star model used for Gaia solutions \citep{2021A&A...649A...2L}. 

Recognizing an optical pair is challenging if both stars belong to the cluster, as their proper motions and parallaxes are inherently similar; however, this configuration is statistically rare. The deviation of reported parallaxes $\varpi \pm \sigma_\varpi$ serves as an additional marker with limited reliability, as this value can be excessive for binaries with confirmed orbital motion \citep{2022MNRAS.517.2925C}:
\begin{equation}
   \Delta \varpi / \sigma =  \frac{|\varpi_A-\varpi_B| }{\sqrt{\sigma^2_{\varpi_A}+\sigma^2_{\varpi_B}}}
    \label{eq:delta_plx}
\end{equation}  

Among optical pairs, the smallest separation is for WDS 03463+2529 with $\rho=2.1^{\prime\prime}$, based on Gaia DR3. Interestingly, at its discovery by \cite{1972A&AS....6..177C}, $\rho$ was just 0.5 arcsec. The observed proper motion, parallax, and radial velocity of the brighter star ($G_1=9.18^{\rm mag}$) contradict Pleiades membership, while the fainter source ($G_2=11.52^{\rm mag}$) belongs to the cluster.  Assuming stars are uniformly distributed, the probability of this configuration across the entire sample is below 2\%.  For other chance alignments with $\rho<3^{\prime\prime}$, the fainter star has $G_2 \gtrsim 18^{\rm mag}$ (Figure \ref{fig:resolved}). The second smallest $\rho$ for a confirmed optical pair where both stars have $G<15^{\rm mag}$ is 9.0 arcsec. After excluding optical pairs, 26 systems with $\delta \mu< 10$ mas~yr$^{-1}$ remain as bound binaries, though this subsample is strongly biased, particularly for $\rho<2^{\prime\prime}$, due to missing two-parameter solutions.

\subsection{Sources with two-parameter solutions}
\label{two-parameter}

The distinction between genuine binaries and optical pairs is ambiguous for sources with two-parameter solutions, common among Gaia-resolved pairs with low $\rho$. For sources with $G<15^{\rm mag}$, these solutions appear only in $\rho \lesssim 1^{\prime\prime}$ systems (Figure \ref{fig:resolved}), which are likely physical. For fainter companions, the chance of optical projection increases. Given randomly distributed field stars and low projection probability ($P \ll 1$), $P= N \rho^2 R^{-2}$, where $\rho$ is angular separation, and $N$ is star count to the fainter component magnitude in $R=2^{\circ}$ radius \citep{1982ApJ...258..589P}. Random projections mainly affect pairs with faint secondaries and larger separations.  The mass ratio of genuine binaries depends on magnitude difference, and the $P<0.2$\% condition is met by all pairs with two-parameter sources and $q>0.2$, meaning they are likely physical. Among entries, WDS 03483+2513 with $G_1=9.12^{\rm mag}$ is unique, as it lacks a secondary magnitude. Its mass ratio $q\sim 0.55$ determined from speckle observations (Table \ref{tab:speckle}) suggests
$G_2\sim12.3^{\rm mag}$. 

\begin{table*}[h]
    \centering
        \caption{List of resolved Gaia DR3 pairs with $\rho<10^{\prime\prime}$ and $G_1<15^{\rm mag}$. Entries with $\delta \mu<10$ mas~yr$^{-1}$ (Equation \ref{eq:pm}) or $P>0.2$\% (for two-parameter solutions; Section \ref{two-parameter}) are selected.  First column:  WDS designation, asterisk (*) marks cases when identifier is newly introduced. The following columns: Gaia DR3 source designation, $G$ magnitude, and \textcolor{black}{mass} for primary and secondary component; angular separation; relative proper motion (Equation \ref{eq:pm}); parallax deviation (Equation~\ref{eq:delta_plx}); projection probability; estimated mass ratio (Section \ref{sample}). The following notes have an associated entry in Section \ref{Membership} or \ref{Multiples} of the Appendix. C: multiplicity is suspected from color-magnitude diagram analysis; F: possible cluster non-member; R: system has another speckle-resolved companion with $\rho<1^{\prime\prime}$ (Table \ref{tab:speckle}); S: spectroscopic binarity related to unresolved companion. \textcolor{black}{Estimated masses are isochrone-based (Section \ref{Mass}) and account for known multiplicity; values are less credible for $M \lesssim 0.5~M_\odot$ stars.}} 
     \resizebox{1 \textwidth}{!}{\hspace{-3cm}\begin{tabular}{lcccccccccccc}
        \hline
        &\multicolumn{3}{c}{Primary component} & \multicolumn{3}{c}{Secondary component} & \multicolumn{6}{c}{System}\\ \hline
       \multicolumn{2}{c}{Designation} &$G_1$&$m_1$&Designation&$G_2$&$m_2$ & $\rho$ & $\delta \mu$& $\Delta \varpi / \sigma_{1,2}$&P &Mass ratio&Notes\\
WDS&Gaia DR3&mag&$M_\odot$&Gaia DR3&mag&$M_\odot$&arcsec&mas/yr&&\%&q& \\ \hline
03459+2433&66798496781121792&5.752&$3.47\pm0.15$&66798526845337344&13.833&$0.58\pm0.01$&8.84&0.93&0.57&0.43&$0.168\pm0.004$& \\ 
03500+2351&66507469798631808&6.806&$2.42\pm0.12$&66507469798631936&10.005&$1.15\pm0.04$&3.28&3.48&0.55&0.00 &$0.48\pm 0.03$&S\\ 
03458+2309&64956127609464320&6.895&$2.45\pm0.12$&64956123313498368&10.127&$1.13\pm0.03$&3.61&1.50&0.35&0.01&$0.460\pm0.010$&\\ 
03509+2358*&66486510358371072&6.930&$2.16\pm0.26$&66486510354964864&15.961&$0.37\pm0.02$&2.78&&&0.16&$0.17\pm0.02$&C \\ 
03456+2420&65282716922610944&7.249&$2.18\pm0.11$&65282716920396160&15.446&$0.43\pm0.02$&5.26&6.22&3.07&0.42&$0.198\pm0.001$&R \\ 
03474+2355&65207709611941376&7.287&$2.16\pm0.11$&65207709613871744&9.739&$1.09\pm0.03$&6.27&2.24&0.56&0.01&$0.50\pm0.03$&S \\ 
03463+2411&65230764996027776&8.229&$1.66\pm0.06$&65230764998207232&9.409&$1.29\pm0.04$&1.06&6.38&0.07&0.00&$0.781 \pm 0.005$& \\ 
03487+2316&64933759417769984&8.337&$1.62\pm0.06$&64933759417767424&17.389&$0.20\pm0.02$&5.01&3.29&0.89&1.14&$0.126 \pm 0.004$&R \\ 
03435+2244&64837242912474624&8.885&$1.44\pm0.05$&64837139834697984&9.766&$1.21\pm0.04$&1.71&1.94&0.81&0.00&$0.840 \pm 0.003$& \\ 
03444+2408&65272821318002560&8.972&$1.40\pm0.05$&65272817023559040&14.206&$0.56\pm0.01$&3.84&9.05&1.05&0.10&$0.40 \pm 0.02$&R \\ 
03483+2513&66939848447027584&9.118&$1.37\pm0.04$&66939852742083328&$\sim 12.3$&$0.76\pm0.05$&1.04&&&0.00&$0.55\pm0.03$&\\ 
03447+2449&69811948914407168&9.715&$1.22\pm0.04$&69811948914509056&15.963&$0.37\pm0.02$&2.76&&&0.16&$0.300 \pm 0.009$ & \\ 
03500+2351&66507469798631936&10.005&$1.15\pm0.04$&66507469798632320&10.235&$1.10\pm0.03$&6.83&4.60&1.14&0.02&$0.95\pm0.03$&R,S\\ 
03434+2314&65063707949772544&10.358&$1.07\pm0.03$&65063707949772672&15.693&$0.40\pm0.02$&3.62&0.85&1.15&0.23&$0.38 \pm 0.03$&R \\ 
03442+2406&65249250535404928&10.878&$0.98\pm0.03$&65249250537488128&11.056&$0.95\pm0.03$&1.84&3.11&3.98&0.00&$0.968\pm0.002$& \\ 
03453+2517&69864313155605120&11.010&$0.95\pm0.03$&69864313154046592&17.267&$0.21\pm0.02$&3.9&2.17&0.48&0.65&$0.22\pm0.02$&S \\ 
03441+2402&65247704349267584&11.167&$0.93\pm0.03$&65248460263511552&11.860&$0.82\pm0.02$&5.69&1.46&1.71&0.05&$0.88\pm0.03$ & R \\ 
03494+2456&66873435368674944&11.242&$0.92\pm0.03$&66873431072788608&11.298&$0.91\pm0.03$&0.28&&&0.00&$0.990 \pm 0.002$ & \\ 
03433+2227*&64814045795106944&11.794&$0.83\pm0.02$&64814050089186048&12.052&$0.80\pm0.02$&0.54&&&0.00&$0.956 \pm 0.002$ & \\ 
03454+2326&65163797868281088&11.943&$0.81\pm0.02$&65163797866480768&15.539&$0.42\pm0.02$&3.37&1.84&0.98&0.18&$0.520 \pm 0.012$ & \\ 
03461+2452&66814302260735104&11.950&$0.81\pm0.02$&66814302258090880&13.523&$0.61\pm0.01$&0.86&&&0.00&$0.752 \pm 0.003$ & \\ 
03473+2344&65010239900405504&11.995&$0.80\pm0.02$&65010244195009024&16.502&$0.29\pm0.02$&1.16&0.97&1.43&0.04&$0.364\pm0.015$ & \\ 
03541+2420&66584332530000512&12.185&$0.78\pm0.02$&66584332532300288&14.897&$0.50\pm0.02$&1.06&&&0.01&$0.640\pm 0.003$ & \\  
03466+2421&66733552578791296&12.441&$0.74\pm0.02$&66733556873061120&12.744&$0.71\pm0.02$&0.89&6.05&1.22&0.00&$0.959\pm 0.005$ & \\  
03532+2356*&66474450090138496&12.567&$0.73\pm0.02$&66474450090139008&15.296&$0.45\pm0.02$&3.71&1.87&0.73&0.19&$0.620\pm0.015$ & F \\ 
03445+2451&69823940463098752&13.281&$0.64\pm0.03$&69823940463098112&18.244&$0.15\pm0.01$&4.62&1.21&0.02&1.46&$0.231\pm 0.006$ & \\ 
03421+2443*&68364544935515392&13.517&$0.61\pm0.02$&68364544933829376&13.593&$0.60\pm0.02$&1.34&1.42&1.63&0.01&$0.989\pm0.003$ &  \\ 
03473+2403&65212691775922048&13.563&$0.61\pm0.02$&65212691773969280&14.674&$0.52\pm0.01$&1.27&0.60&0.65&0.02&$0.856\pm0.004$& \\ 
03486+2246*&64131150292139648&13.846&$0.58\pm0.01$&64131150289171712&17.224&$0.22\pm0.02$&0.8&&&0.03&$0.374 \pm 0.017$ & \\ 
03427+2412&68267680536205440&13.849&$0.58\pm0.01$&68267684833497728&16.354&$0.24\pm0.02$&1.71&&&0.08&$0.41\pm0.03$ & R\\ 
03475+2223*&64021920682268416&13.977&$0.57\pm0.01$&64021920680680576&16.696&$0.27\pm0.02$&1.01&&&0.03&$0.472\pm0.022$ & \\ 
03461+2423&66781316909399424&14.012&$0.57\pm0.01$&66781316912634880&18.479&$0.13\pm0.01$&1.16&&&0.10&$0.237\pm0.008$ & \\ 
03436+2414&65266494828710400&14.179&$0.56\pm0.01$&65266499126062080&14.437&$0.54\pm0.01$&1.84&4.47&2.74&0.03&$0.964\pm0.004$ & \\ 
03462+2440&66801623514684416&14.265&$0.55\pm0.01$&66801623517294848&16.358&$0.31\pm0.02$&1.38&5.21&1.24&0.05&$0.563\pm0.028$ & \\ 
03445+2353*&65241313435901568&14.321&$0.55\pm0.01$&65241313437941504&14.405&$0.54\pm0.01$&1.47&1.11&0.19&0.02&$0.988\pm0.002$ & \\ 
03502+2421*&66555573432261376&14.463&$0.54\pm0.02$&66555573432261120&15.331&$0.45\pm0.02$&1.95&2.45&0.01&0.05&$0.838\pm0.019$ & \\ 
03459+2552&70035351636365824&14.520&$0.53\pm0.01$&70035355933666304&15.656&$0.32\pm0.02$&1.58&2.91&1.16&0.04 &$0.60\pm0.04$&R\\ 
03508+2240*&64094037476647040&14.974&$0.49\pm0.02$&64094041774819712&17.591&$0.19\pm0.01$&1.20&&&0.07&$0.383\pm0.010$& \\ 
    \end{tabular}}
    \label{tab:resolved}
\end{table*}

\begin{table*}[h]
    \centering
        \caption{Speckle-resolved systems with $\rho<1.05^{\prime\prime}$ (Section \ref{results}). The full observational log is available in Table \ref{tab:binarity_observations} (Appendix).
An asterisk (*) marks cases where the WDS name is newly introduced. Notes. B: source is blended; $G$-band magnitude represents the total flux of components (Section \ref{resolved_types}). D: a distant companion is present in Gaia (Table \ref{tab:resolved}). F: possibly a cluster non-member (Section \ref{Membership}, Appendix). G: this pair is Gaia-resolved (Table \ref{tab:resolved}). N: the companion is previously unreported (Section \ref{discoveries}). S: spectroscopic binarity related to unresolved companion (Section \ref{Multiples}, Appendix). T: three resolved components within 1 arcsec.}
 \resizebox{0.93 \textwidth}{!}{\hspace{-2.7cm}\begin{tabular}{ccccccccccccc}
        \hline
       \multicolumn{2}{c}{Designation} & $G$ &Date&$\rho$&Band&  Contrast&\multicolumn{2}{c}{Mass}& \multicolumn{3}{c}{Mass ratio} &Notes \\
                \hline
                WDS&Gaia DR3 & mag & yr& arcsec&&mag&$m_1,~ M_\odot$&$m_2,~ M_\odot$&$q_{\rm min}$&$q$&$q_{\rm max}$& \\
                \hline
03463+2357&65205373152172032&4.173&2023.846&0.27&$\hat{z}$&4.8&$4.79\pm0.09$&$1.21\pm0.09$ &0.24&0.25&0.27&B\\
03456+2420&65282716922610944&7.249&2023.846&0.26&$\hat{z}$&1.2&$2.19\pm0.11$&$1.42\pm0.11$&0.61&	0.65&	0.70&D \\
03482+2419&66724451545088128&8.214&2024.163&0.07&$\hat{y}$&1.6&$1.58\pm0.06$&$1.15\pm0.06$&0.69	&0.73	&0.76&B\\
03487+2316&64933759417769984&8.337&2023.920&0.85&$\hat{z}$&4.3&$1.62\pm0.06$&$0.59\pm0.04$&0.35&	0.37&	0.39&D,N\\
03471+2449&66832993955739776&8.479&2024.240&0.62&$\hat{z}$&1.2&$1.57\pm0.05$&$1.16\pm0.07$&0.70&	0.74&	0.78&\\
03475+2406&66715174415764736&8.624&2023.846&0.15&$\hat{z}$&1.1&$1.35\pm0.04$&$1.16\pm0.07$&0.80&0.86&0.92&S\\
03444+2408&65272821318002560&8.972&2023.846&0.08&$\hat{z}$&3.2&$1.40\pm0.05$&$0.71\pm0.05$&0.47&	0.50&	0.53&B,D,N\\
03486+2411&66718610389577088&9.012&2024.163&0.25&$\hat{z}$&1.1&$1.40\pm0.04$&$1.09\pm0.07$&0.74&	0.78&	0.82&\\
03483+2513&66939848447027584&9.118&2024.240&1.01&$\hat{z}$&2.7&$1.37\pm0.04$&$0.76\pm0.05$&0.53&	0.55&	0.59&G\\
03457+2454&69816346960886784&9.424&2023.846&0.07&$\hat{y}$&0.2&$1.29\pm0.04$&$1.22\pm0.06$&0.91&	0.95&	0.99&\\
03481+2409&66713044112564864&9.441&2023.920&0.53&$\hat{y}$&1.3&$1.29\pm0.04$&$1.01\pm0.05$&0.75	&0.78	&0.82&\\
03392+2428&68254245878512384&10.226&2023.810&0.83&$I_c$&5.8&$1.11\pm0.03$&$0.24\pm0.04$&0.18&	0.21&	0.25&N\\
03500+2351&66507469798632320&10.235&2024.002&0.53&$I_c$&2.6&$1.10\pm0.03$&$0.62\pm0.05$&0.53&0.57&0.62&S\\
03434+2314&65063707949772544&10.358&2024.590&0.11&$I_c$&2.6&$1.07\pm0.03$&$0.61\pm0.04$&0.54&	0.57&	0.61&B,D,N\\
03491+2347&66503449709270400&10.431&2024.002&0.04&$I_c$&0.7&$0.99\pm0.03$&$0.85\pm0.04$&0.81&	0.86&	0.91&B\\
03488+2416&66720946851771904&10.706&2024.240&0.10&$I_c$&0.4&$0.92\pm0.03$&$0.82\pm0.06$&0.79&0.89&0.96&B,S\\
03438+2313&65063089472971776&10.774&2024.631&0.42&$I_c$&0.8&$1.00\pm0.03$&$0.84\pm0.05$&0.80&	0.84&	0.89&N\\
03438+2415&65289279632597760&11.109&2023.810&0.19&$I_c$&3.7&$0.94\pm0.03$&$0.41\pm0.05$&0.39&0.44 &	0.48&B,N\\
03436+2346&65233788655261568&11.202&2023.996&0.20&$I_c$&3.7&$0.92\pm0.03$&$0.40\pm0.05$&0.39&	0.44&	0.48&B,N\\
03494+2456&66873435368674944&11.242&2024.240&0.25&$I_c$&0.2&$0.92\pm0.03$&$0.87\pm0.05$&0.90&	0.95&	0.99&G\\
03509+2350&66458610251617536&11.408&2024.002&0.26&$I_c$&0.7&$0.89\pm0.03$&$0.77\pm0.05$&0.82&0.86&0.91\\ 
03447+2553&69945814454871680&11.438&2024.637&0.58&$I_c$&4.5&$0.89\pm0.03$&$0.26\pm0.04$&0.26&	0.29&	0.33&N,T\\
03447+2553&69945814454871680&11.438&2024.637&0.66&$I_c$&4.6&$0.89\pm0.03$&$0.25\pm0.03$&0.25&	0.28&	0.31&N,T\\ 
03405+2429&68334235349446528&11.510&2023.810&0.18&$I_c$&3.1&$0.87\pm0.02$&$0.44\pm0.04$&0.47&0.51&	0.55&B,N\\ 
03520+2440&66665696393227904&11.729&2023.996&0.11&$I_c$&0.2&$0.84\pm0.02$&$0.80\pm0.04$&0.90&	0.95&	0.98&T\\
03520+2440&66665696393227904&11.729&2023.996&0.42&$I_c$&1.5&$0.84\pm0.02$&$0.60\pm0.04$&0.68&	0.71&	0.75&T\\ 
03433+2227*&64814045795106944&11.794&2023.843&0.55&$I_c$&0.3&$0.83\pm0.02$&$0.78\pm0.04$&0.89	&0.93&	0.98&G\\
03441+2402&65248460263511552&11.860&2024.002&0.25&$I_c$&2.8&$0.82\pm0.02$&$0.44\pm0.04$&0.50&0.54&0.59&B,D,N\\
03457+2345&65178568258674176&11.926&2024.002&0.53&$I_c$&2.7&$0.81\pm0.02$&$0.46\pm0.04$&0.52&	0.57&	0.61&\\
03461+2452&66814302260735104&11.950&2024.002&0.85&$I_c$&1.3&$0.81\pm0.02$&$0.61\pm0.04$&0.72&	0.75&	0.81&G\\
03507+2524&66957612431742080&12.017&2024.240&0.17&$I_c$&0.3&$0.73\pm0.02$&$0.69\pm0.03$&0.88&0.94&0.98&B,N\\
03434+2500&69876506565909632&12.238&2024.002&0.75&$I_c$&2.5&$0.77\pm0.02$&$0.44\pm0.04$&0.52&0.58&0.62&\\
03466+2421&66733552578791296&12.441&2024.240&0.92&$I_c$&0.6&$0.74\pm0.02$&$0.66\pm0.05$&0.82&0.89&	0.95&G\\
03445+2410&65273027476967680&12.944&2023.844&0.44&$I_c$&2.2&$0.69\pm0.02$&$0.40\pm0.04$&0.53&	0.58	&0.64&\\
03492+2211&63958561324922240&12.994&2024.160&0.13&$I_c$&1.9&$0.67\pm0.03$&$0.43\pm0.04$&0.59&	0.64&	0.70&B,N\\
03471+2343&65010175477867776&13.318&2024.161&0.63&$I_c$&0.4&$0.63\pm0.03$&$0.59\pm0.03$ &0.88&	0.93&	0.97&\\ 
03467+2456&66837495084120320&13.436&2023.844&0.46&$I_c$&3.1&$0.62\pm0.02$&$0.24\pm0.03$&0.34	&0.38&	0.43&T\\
03467+2456&66837495084120320&13.436&2023.844&0.42&$I_c$&3.1&$0.62\pm0.02$&$0.24\pm0.03$&0.34	&0.38&	0.43&N,T\\ 
03419+2327&65131121756446592&13.488&2024.002&0.89&$I_c$&1.7&$0.61\pm0.02$&$0.42\pm0.04$&0.62&0.68&0.74&	\\
03465+2407&65226401312036864&13.536&2023.844&0.29&$I_c$&2.3&$0.61\pm0.02$&$0.33\pm0.04$&0.47&	0.54&	0.60&\\
03496+2327&64944170420647296&13.602&2023.844&0.65&$I_c$&2.6&$0.60\pm0.01$&$0.28\pm0.04$&0.41	&0.47&	0.53&\\
03492+2333&64949427460614016&13.676&2023.844&0.61&$I_c$&2.3&$0.60\pm0.01$&$0.32\pm0.04$&0.47&0.53&0.60&\\
03486+2246*&64131150292139648&13.846&2024.161&0.80&$I_c$&3.1&$0.58\pm0.01$&$0.21\pm0.03$&0.32&	0.36&	0.40&G\\
03475+2223*&64021920682268416&13.977&2024.161&1.01&$I_c$&2.6&$0.57\pm0.01$&$0.24\pm0.03$&0.38&	0.42&	0.48&G\\
03484+2358&66521213693847808&14.002&2024.160&0.10&$I_c$&1.0&$0.57\pm0.01$&$0.46\pm0.04$&0.75	&0.81&	0.87&N\\
03534+2324&65690498298435840&14.073&2024.628&0.09&$I_c$&1.6&$0.55\pm0.01$&$0.35\pm0.04$&0.57&	0.64&0.72&B,N\\
03404+2249&64664997545142784&14.270&2024.002&0.23&$I_c$&0.3&$0.55\pm0.01$&$0.52\pm0.03$&0.89&	0.94	&0.98&N\\
03478+2233&64033774791971328&14.514&2024.161&0.21&$I_c$&0.7&$0.53\pm0.01$&$0.45\pm0.04$&0.77&	0.85&	0.91&F,N\\
03493+2404&66525783539163264&14.588&2024.629&0.15&$I_c$&0.3&$0.53\pm0.01$&$0.49\pm0.03$&0.88&0.94&0.98&F,N\\
03437+2434&68301250000967680&14.652&2023.844&0.76&$I_c$&1.8&$0.52\pm0.01$&$0.28\pm0.04$&0.47	&0.54&	0.61&\\
03476+2342&65007220540365184&14.656&2024.628&0.15&$I_c$&0.4&$0.52\pm0.01$&$0.47\pm0.04$&0.83&	0.91&0.97&N\\
03491+2344&66500013736315264&14.776&2023.844&0.45&$I_c$&0.7&$0.51\pm0.01$&$0.42\pm0.04$&0.76&	0.83&	0.91\\
03502+2400&66511382510098560&14.779&2023.844&0.61&$I_c$&0.8&$0.51\pm0.01$&$0.40\pm0.04$&0.72&0.80&0.87\\
03427+2412&68267684833497728&16.354&2023.844&0.21&$I_c$&0.2&$0.24\pm0.02$&$0.21\pm0.02$&0.80	&0.90&	0.97&D,N\\
    \end{tabular}}
    \label{tab:speckle}
\end{table*}

\section{Observations and data reduction}
\label{observations}
\subsection{Speckle interferometry}
Speckle interferometric observations were conducted from October 2023 to October 2024 with the speckle polarimeter on the 2.5 m telescope \citep{2020gbar.conf..127S} at the Caucasian Observatory of Sternberg Astronomical Institute, Moscow State University (SAI MSU). The instrument features a high--speed, low--noise CMOS detector, the Hamamatsu ORCA–quest \citep{2023AstBu..78..234S}. Median seeing, estimated as the stellar image's FWHM in average frames, was 0.8 arcsec. An atmospheric dispersion compensator was active during observations. For objects with $G<12.5^{{\rm mag}}$, 4000 frames with 30 ms exposure were obtained. For $G\geq12.5^{{\rm mag}}$ sources, 2000 frames with 60 ms exposure were acquired.  Stars with $G<9.5^{{\rm mag}}$ were observed in 550 and 880 nm filters (50 and 70 nm half-widths), for fainter objects $I_c$ band was used (\textcolor{black}{Figure \ref{fig:filters})}. A few objects were also observed in the $R_c$ band. The angular scale of $20.56$~mas~px$^{-1}$, with frame dimensions of the registered datacubes at $512 \times 256$~px, provides a $10.5 \times 5.3$ arcsec field.

Speckle interferometry \citep{1970A&A.....6...85L} is based on the assumption of isoplanatism, which states that the point spread function (PSF) remains constant across the field of view. In practice, light from 
resolved sources travels along slightly different paths through the turbulent atmosphere, leading to variations in wavefront distortions \citep{1981PrOpt..19..281R}. The isoplanatic angle (the angular separation at which PSF differences are minimal) depends on atmospheric conditions and the observational wavelengths. The contrast estimations are also affected by the instrumental CMOS-specific rolling shutter effect, which is a slight temporal lag between sensor row readouts. 
In our case, the effects of anisoplanatism and rolling shutter begin to impact contrast estimations at angular distances larger than 1 arcsec; therefore, we rely on Gaia data for $\rho>1^{\prime\prime}$ binaries (Section \ref{Gaia-resolved}).  

Each series was reduced and processed using the technique from \cite{2023AstBu..78..234S} to compute the average power spectrum $\langle{|\widetilde{I}|}^2\rangle$. Besides averaging over the entire series, we calculated the bootstrapped \citep{Efron1993} average power spectra $\langle{|\widetilde{I}|}^2\rangle_{B}$, using 30 random bootstrap subsamples of frame indices for averaging, each with length equal to the original series. 

The image $I$ is a convolution of the object intensity
distribution $O$ and the PSF $T$. In the Fourier space, 
$\widetilde{I}(\boldsymbol{f})=\widetilde{O}(\boldsymbol{f})\,\widetilde{T}(\boldsymbol{f})$,
where $\boldsymbol{f}$ is the spatial frequency vector, $\widetilde{T}$ is the optical transfer function, and $\widetilde{O}$ is the object visibility. We estimate the squared modulus of the object visibility ${|\widetilde{O_e}|}^2$, and approximate it with a known model function. The autocorrelation function \textcolor{black}{(ACF)}
is calculated \textcolor{black}{using} the inverse Fourier transform of ${|\widetilde{O_e}|}^2$.

Most observations of binary stars had single stars observed closely in time, allowing  us to estimate the visibility squared modulus, ${|\widetilde{O_e}|}^2$, by normalizing the averaged power spectra, $\langle{|\widetilde{I}|}^2\rangle$, with the reference single star's averaged power spectrum, $\langle{|\widetilde{I}_{ref}|}^2\rangle$. When multiple reference observations were available, we summed their 
averaged power spectra and used the result for normalization. For binaries without close-in-time reference stars, and for single stars, we normalized the averaged power spectra by their azimuthal average. Reference star normalization compensates for telescope jitter, optical aberrations, and residual atmospheric dispersion, resulting in more accurate binarity estimates compared to azimuthal average normalization {\citep{Strakhov2024}}.

We estimate both non-bootstrapped ${|\widetilde{O_e}|}^2$ and bootstrapped ${|\widetilde{O_e}|}^2_{B}$ versions of visibility squared modulus. If reference power spectra were available, each object’s bootstrap realization $\langle{|\widetilde{I}|}^2\rangle_{B}$ was normalized by the corresponding reference $\langle{|\widetilde{I}_{ref}|}^2\rangle_{B}$. Otherwise, the azimuthal average normalization was used for $\langle{|\widetilde{I}|}^2\rangle_{B}$. 

The binarity parameters of each observation (separation $\rho$, position angle $\theta$ and contrast $\epsilon$) were estimated from a binary (or triple) model fit of non-bootstrapped ${|\widetilde{O_e}|}^2$, using \mbox{MATLAB} implementation of the Levenberg–Marquardt algorithm (\verb|nlinfit|). The standard deviation over bootstrapped ${|\widetilde{O_e}|}^2_{B}$ across 30 bootstrap subsamples served as observation weights. Uncertainties were determined by applying the same model fit to each bootstrapped ${|\widetilde{O_e}|}^2_{B}$ and calculating the standard deviation of the resulting bootstrapped parameters.

The speckle interferometry method inherently loses phase information due to the modulus operation in power spectrum calculations. To resolve the arising 180 degree position angle ambiguity, we estimate the phase of the object visibility from the average bispectrum \citep{1983ApOpt..22.4028L,2004PASP..116...65T} using a recursive algorithm. This phase restoration generally recovers the position angle of a component, except for very close or twin-like pairs with $\rho \lesssim 0.1^{\prime\prime}$ or $\epsilon \lesssim 1^{\rm mag}$. 

\textcolor{black}{The detection limits $\epsilon_{lim}(\rho)$ are calculated using a technique from \cite{2023AstBu..78..234S}. Briefly, we estimate a $5\sigma$ level above the average ACF level in the ring at a given $\rho$. A companion with a magnitude difference less than $\epsilon_{lim}(\rho)$ would be reliably detected by us. Table~\ref{tab:detection_limits} contains the detection limits $\epsilon_{lim}(\rho)$ at selected separations within $\rho=0.1$ and 1 arcsec; the contrast curves are also available at \href{https://doi.org/10.5281/zenodo.14252721}{10.5281/zenodo.14252721}.}

\subsection{Photometric accuracy and precision}
\label{errors}

The median contrast uncertainty from bootstrap realizations is $0.03^{\rm mag}$, with 87\% of formal errors below $0.1^{\rm mag}$ for $0.1<\rho<1$ arcsec (resolved triple systems are excluded from the statistics). Nominal errors are underestimated. For 30 binaries with more than one observation in the same filter, the pooled standard deviation is  $0.18^{\rm mag}$. We adopt this number as the contrast measurement precision. Observation adopted for $q$ calculation is chosen manually according to object visibility squared modulus ${|\widetilde{O_e}|}^2$ appearance, taking into account signal-to-noise,  atmospheric and instrumental conditions. 
Pleiades stars show various modes of intrinsic variability due to stellar flares \citep{2021A&A...645A..42I}, pulsations \citep{2023ApJ...946L..10B}, and spot-induced rotational modulation \citep{2016AJ....152..113R}, though instrumental factors likely drive most of the observed discrepancy.

Contrast measured in speckle observations tends to be overestimated compared to Gaia data. While expected for wider pairs due to anisoplanatism, it persists for systems with $\rho \lesssim 0.6^{\prime\prime}$, showing a $0.25^{\rm mag}$ average residual according to figure 6 from \cite{2024AJ....167...56C}. Direct comparison with Gaia magnitudes is complicated by the passband difference.
The $G$ band, with FWHM $\Delta \lambda =455$ nm \citep{2021A&A...649A...3R}, is broader than filters used for speckle observations. Open clusters allow estimation of mass ratios based on flux contrast, thanks to elaborate isochrone models. Then, for binaries with known $q$, we calculate the expected magnitude difference in $G$ passband. Among $\rho<1^{\prime\prime}$ pairs, only five are resolved by Gaia (Section \ref{tab:resolved}), and our estimate of $\Delta G$ based on speckle observations is underestimated by $0.18^{\rm mag}$ on average, with a $0.13^{\rm mag}$ standard deviation.

The infrared adaptive optics observations from \cite{1997A&A...323..139B} cover 7 pairs with $0.25<\rho<1$ arcsec, excluding multiple systems with spectroscopic binarity. Our \textcolor{black}{synthetic $\Delta J$ estimates for these entries, based on $q$ derived with speckle observations,} are overestimated by $0.05^{\rm mag}$ with a $0.11^{\rm mag}$ scatter. Another adaptive optics survey, using $i^\prime$ and LP600 filters,  approximately corresponding to the $I_c$ passband \citep{2018AJ....155...51H}, observed 8 binaries in the $0.45<\rho<1$ arcsec range. Our $\Delta I_c$ contrasts for these binaries are lower by $0.18^{\rm mag}$ with a $0.2^{\rm mag}$ residual. Finally, for WDS~03491+2347 (${\rho \sim 0.04^{\prime\prime}}$, Table \ref{tab:v1282}), we overestimate $\Delta J$ by $0.22^{\rm mag}$. Due to fairly large discrepancies, we tentatively adopt systematic contrast accuracy equal to a precision of $0.18^{\rm mag}$, combining to ${\sqrt{2\cdot0.18^2} \sim 0.25^{\rm mag}}$ uncertainty. For each observation, we add the nominal measurement error, though it is usually negligible. We note that a relatively large error leads to an underestimation of $q$ for twin binaries, as $q \leq 1$ by definition. 

\section{Speckle observations results}
\label{results}

We directly detect 49 pairs and 3 triples with $\rho<1^{\prime\prime}$ (Table \ref{tab:speckle}). The widest speckle-resolved binary star unresolved in Gaia has $\rho=0.89^{\prime\prime}$  (WDS 03419+2327). We also rely on speckle observations for WDS 03483+2513 with $\rho=1.04^{\prime\prime}$, as it lacks a $G_2$ magnitude. For wider pairs, Gaia exceeds our measurements both in accuracy and sensitivity. We confidently detect five pairs closer than 0.1 arcsec, including WDS 03491+2347 with $\rho=0.04^{\prime\prime}$. The angular separation and estimated mass ratio of resolved pairs are shown in Figure \ref{fig:results}. 

\begin{figure*}
    \centering
    \includegraphics[width=1.0\linewidth]{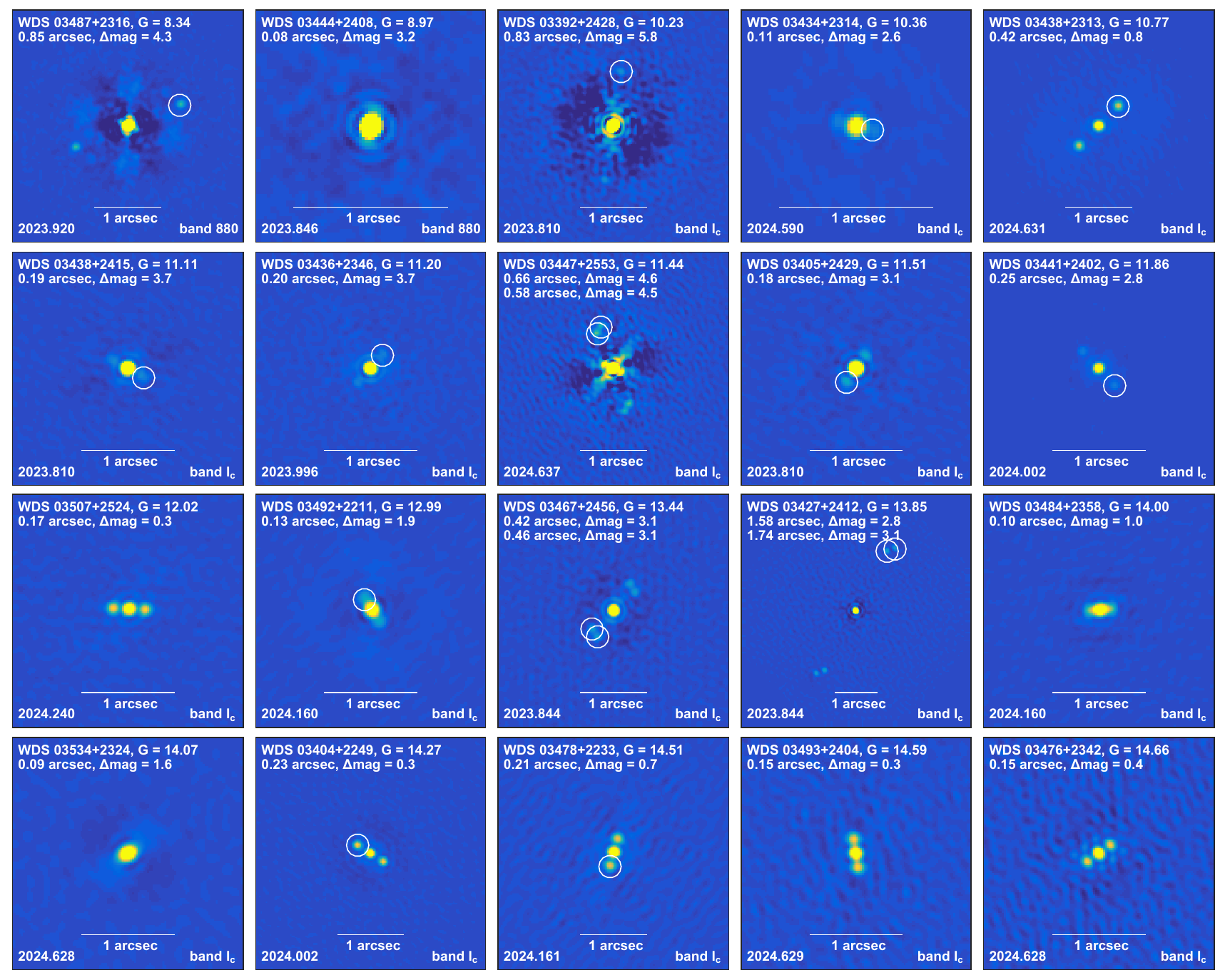}  
    \caption{The images of the speckle autocorrelation function for the discovered systems (Section \ref{discoveries}). Circles indicate the  recovered position angle; North is up, East is left. Images of all resolved systems and contrast curves are available at \href{https://doi.org/10.5281/zenodo.14252721}{Zenodo}.} 
    \label{fig:icons}
\end{figure*}

\begin{figure*}
    \centering
    \begin{minipage}{\linewidth}
    \includegraphics[width=\linewidth]{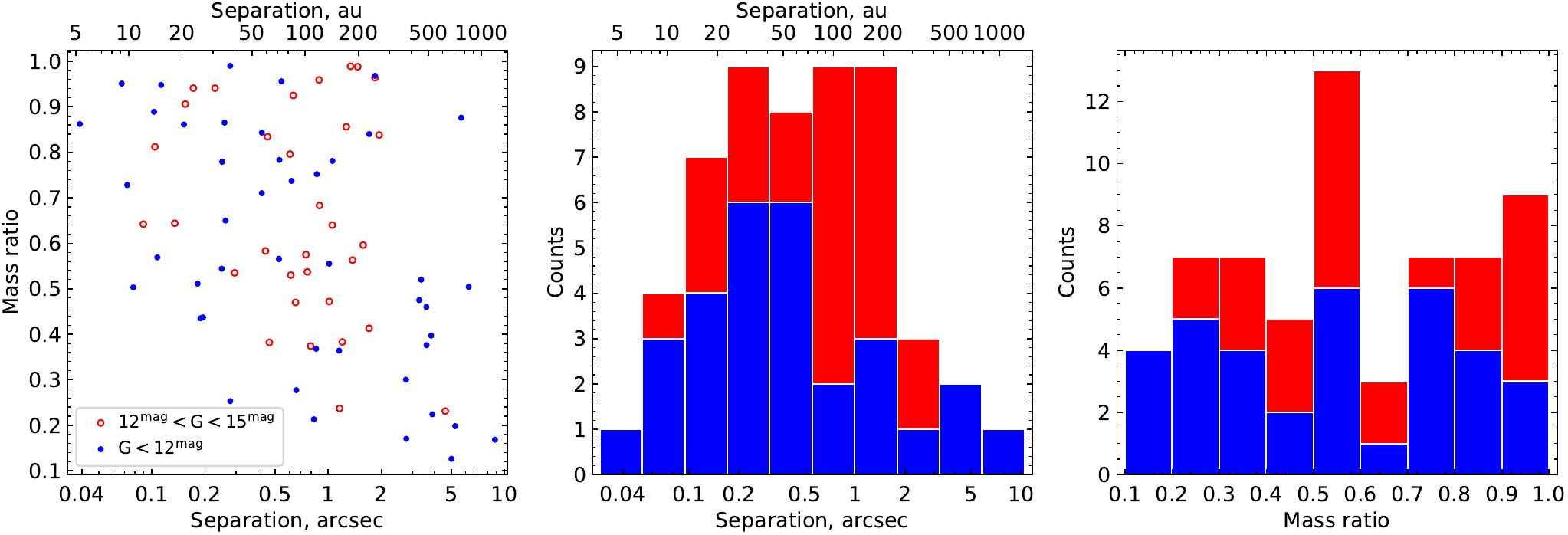}
    \end{minipage}
    \caption{Mass ratio and separation of the resolved binaries in the Pleiades. \textcolor{blue}{\bf Blue} is used for binaries with primary magnitude $G_1<12^{\rm mag}$, and \textcolor{red}{\bf red} for $12^{\rm mag}<G_1<15^{\rm mag}$ .  In the \textbf{middle panel}, only $q>0.5$ systems are considered. For the \textbf{right panel}, $\rho>0.2^{\prime\prime}$ pairs are selected to eliminate incompleteness. While it is tempting to see hints of bimodality in $q$ distribution, we beware of over-interpretation, as the sample size is small, and the accuracy of speckle measurements is limited.}
    \label{fig:results}
\end{figure*}

\subsection{Previously unreported components}
\label{discoveries}

We report the detection of 21 companions in 20 systems for the first time (Figure \ref{fig:icons}); Gaia-resolved pairs are considered as known. The discovered systems were observed at least twice for confirmation. Two new components were revealed in the triple system WDS 03447+2553 (${G=11.44^{\rm mag}}$). A third component was found for six systems previously known as binary. The primary component was resolved into a close pair for WDS 03487+2316 (${G_1=8.34^{\rm mag}}$), WDS~03444+2408 ($G_1=8.97^{\rm mag}$), and WDS~03434+2314 (${G_1=10.36^{\rm mag}}$). Conversely, in the cases of WDS 03441+2402 ($G_1=11.17$, ${G_2=11.86^{\rm mag}}$) and WDS 03427+2412 ($G_1=13.85$, $G_2=16.35^{\rm mag}$) the fainter Gaia source was resolved. For WDS~03467+2456 ($G=13.44^{\rm mag}$), unresolved in Gaia but discovered as binary by \cite{2018AJ....155...51H}, we reveal that the fainter component itself is a pair. More details for triple stars are provided in Section \ref{Multiples} of the Appendix. Finally, in 13 cases marked with the note ‘N’ in Table \ref{tab:speckle}, we detected previously unresolved stars as binary. 

\subsection{Merope, 23 Tau}
\label{Merope}
Merope (23 Tau, WDS 03463+2357, ${G=4.17^{\rm mag}}$) is the only bright star ($G$ or $V<6^{\rm mag}$) resolved by our speckle observations. The companion was discovered in 2002 by \cite{2011MNRAS.415..854D} with the 3.63 m Advanced Electro Optical System Telescope. The measured contrast $\Delta I_c = 4 \pm 0.4^{\rm mag}$ complies with $0.28<q<0.36$ range. However, more recent observations failed to resolve Merope: \cite{2020AJ....159..233H} reported a non-detection at the 5$\sigma$ level with 4.3 m Lowell Discovery Telescope up to 4.12 magnitude difference at $\rho=0.2^{\prime\prime}$ with 880 nm filter. \cite{2021ApJS..257...69H} after observations with Navy Precision Optical Interferometer also concluded that 23 Tau is likely a single star. Our detection with $4.8 \pm 0.2^{\rm mag}$ contrast in the 880 nm band shows a 56 degree change of position angle over 21.5 years, while the separation slightly increased from 0.25 to 0.28 arcsec. The estimated mass ratio is $0.25\pm 0.02$. The dynamical mass $6.00\pm0.14 \ M_\odot$ roughly corresponds to 90 yr orbital period assuming circular face-on orbit. Notably, \cite{2022A&A...657A...7K} reported $6.48 \pm 0.25$ km~s$^{-1}$ Hipparcos-Gaia tangential velocity anomaly  for Merope.  

\section{Multiplicity statistics}
\label{statistics}
\subsection{Sample completeness}
\label{completeness}

In this paper, we restrict our analysis to resolved systems. The unresolved multiplicity is considered only when it affects the mass ratio calculation for resolved pairs in multiple systems (Section \ref{Multiples}, Appendix). The lists of Gaia- and speckle-resolved pairs (Tables \ref{tab:resolved} and \ref{tab:speckle}) contain 38 and 54 entries, respectively, with 7 systems having $\rho<1.05^{\prime\prime}$ appearing in both. We exclude three binaries that are not confirmed cluster members (Appendix, Section \ref{Membership}) and the Gaia-resolved pair WDS~03500+2351 with $\rho=6.8^{\prime\prime}$, whose components are both secondaries in the sextuple system (Table \ref{tab:hd23964}). This leaves us with 81 pairs in 61 systems. The ten systems are  resolved triples, both inner and outer pairs included. Three inner binaries are associated with $G_2<15^{\rm mag}$ companions and thus fall outside the threshold. In total, the 78 pairs shown in Figure \ref{fig:results} remain for analysis.  

As follows from Figure \ref{fig:Gaia}, pairs with ${G_2-G_1<4^{\rm mag}}$ are resolved in Gaia for $\rho>1^{\prime\prime}$, ensuring a ${q>0.5}$ completeness. This constraint extends at least to ${\Delta G< 6^{\rm mag}}$ and $q>0.35$ at 2 arcsec. Chance alignments among the $\rho<10^{\prime\prime}$ entries are recognized thanks to astrometric data (Figure \ref{fig:resolved}). 
The atmospheric and instrumental conditions affect the ability to resolve double stars in speckle observations. For unresolved sources, the detection limit is estimated; the sample is expected to be complete for $q>0.6$ pairs at $\rho>0.2^{\prime\prime}$, with the incompleteness level for $q>0.5$ pairs as low as 3\%. With a few caveats, we anticipate a $q>0.5$ completeness in the 0.2 -- 10 arcsec range; 39 entries meet these criteria.
\vspace{1mm}

\subsection{\textcolor{black}{Wide binary statistics}}
\label{Wide}
We explored 204 and 205 probable Pleiades members in the bright ($G<12^{\rm mag}$, $m\gtrsim0.8 M_\odot$) and faint ($12^{\rm mag} < G < 15^{\rm mag}$, $0.5 \sim 0.8\ M_\odot$ range) samples respectively. 
Nine and eleven Gaia sources are secondary components in their systems (Section \ref{Gaia-resolved}), which leaves us with 195 and 194 entries in the two groups. These numbers serve as the denominator of the binary fraction $f=N_{\rm b} / N$,
where the numerator $N_{\rm b}$ counts binary and multiple systems, each system contributes as one entry.

We count $14^{+2}_{-0}$ pairs with $q>0.6$ and $19^{+0}_{-1}$ with ${q>0.5}$ in the bright sample. The corresponding numbers for the equally sized faint sample are consistent: $12^{+5}_{-0}$ and $19^{+1}_{-3}$ entries. Overall, the estimated binary fraction is ${f=6.7\%^{+1.8}_{-0}}$ and $9.8\%^{+0.3}_{-1.0}$  for $q>0.6$ and $q>0.5$, respectively, within the 27 -- 1350 au range. Including the $\rho<0.2^{\prime\prime}$ resolved pairs with $q>0.5$, the bias-affected value of binary fraction increases to 13\%. 

\textcolor{black}{Low sample size limits our capacity to put firm constraints on separation and mass ratio distributions. The latter,  $f(q)\sim q^\gamma$, is compatible (p-value above 0.1 in the Anderson-Darling test) with $\gamma=-0.2\pm1.2$ within the $27<s<1350$ au and $q>0.5$ domain, and is thus largely inconclusive. The projected separation distribution agrees with a truncated power law $f(s) \sim s^\beta$, $\beta=-1.43\pm0.17$,  ruling out a flat distribution in logarithmic scale. However,  $\beta=-1$ is actually the best-fit value for the 27 -- 300 au range. A sharp drop in binary frequency appears for wider systems (Figure \ref{fig:results}): we identify seven pairs with $q>0.35$ ($f=1.8\%$) within the 2--10 arcsec range, only one of which has $q>0.55$.} 

\subsection{Comparison with other works} 
\label{Comparison}

\textcolor{black}{The interpretation of multiplicity statistics for field binaries seems to be more challenging than in young single-age populations, such as the Pleiades. For older stars, a number of components have experienced significant mass loss and have evolved into degenerate objects such as white dwarfs. We consider stars of $1\pm0.25~M_\odot$ mass in the 25~pc solar neighborhood, referring to figure 28 from \cite{2017ApJS..230...15M}, which shows a log P -- q plot based on \cite{2010ApJS..190....1R} data. For a system with a total mass of $1.5M_\odot$ in the Pleiades, the 0.2 -- 10 arcsec range corresponds to $4.62 < \log P \ {\rm (days)} < 7.17$, assuming circular face-on orbits. We count 44 pairs with $q>0.5$, and the total number of solar-type primaries that have been the most massive components throughout the lifetime of the system is estimated as 404, yielding a 10.9\% fraction. This value agrees well with the obtained Pleiades value of 9.8\%. However, this comparison involved numerous assumptions and should be taken with a grain of salt.}

\textcolor{black}{A deficit of wide systems in open clusters was addressed by \cite{2020MNRAS.496.5176D}, who found a significant shortage of companions at 300 -- 3000 au separations relative to the field ($f=2.3$\% versus 7.8\%). It may be a sign of a dense star formation region. At present, the boundary between soft and hard binary regimes \citep{2023ApJ...955..134R} in the Pleiades is on the order of 1500~au, but dynamical destruction could have affected closer systems since the cluster was probably more compact in the past. Similarly to the previous paragraph, we estimate $f=22/404= 5.4\%$ for the field binary fraction within $6.12 < \log P \ {\rm (days)} < 7.17$ and $q>0.35$, which is three times larger than $f=1.8$\% for the corresponding 2~--~10 arcsec range in the Pleiades.}

\textcolor{black}{Relative lack of large $q$ pairs at 200 -- 500~au was noticed in the young Upper Scorpius association \citep{2020AJ....159...15T}. 
With an adopted distance of 140 pc, it is just 5 pc further away than Pleiades, allowing direct comparison. Its fraction of $q>0.35$ pairs within $270<s<1350$ au range, $f=25/572=4.4\%$ is compatible with the field. The relative proportion of $q>0.7$ to $q>0.35$ systems is 10/22=0.45 in the field, 10/25=0.4 in Scorpius, and just 1/7=0.15 in Pleiades. We beware of ambitious conclusions given low statistics counts. Large mass ratio is an indicator of components interrelation during their formation and early evolution, therefore twin stars often appear in close systems \citep{2019MNRAS.489.5822E}. In a dense environment, pairs of common origin within the same circumbinary disk could be disrupted and replaced by random companions captured in a three-body interaction \citep{2024MNRAS.531..739G}}.  

\subsection{\textcolor{black}{Extrapolated binary fraction}}
\label{fraction}

The explored 27 -- 1350 au range accounts for approximately  1/2.5 fraction of the field binaries population \citep{2023ASPC..534..275O}; therefore, we roughly extrapolate the total binary fraction to be around 25\% for $q>0.5$ systems, and 17\% for $q>0.6$. Given bright and faint samples show similar binary frequency, \textcolor{black}{our results for the stars of different mass are compatible. This is rather unexpected, as \cite{2021AJ....162..264J} showed a steady increase of binary frequency with primary mass in the Pleiades. However, \cite{2020AJ....159...15T} found that in the Upper Scorpius region binary fraction is invariant of primary mass unlike anticipated for field objects. Moreover, \cite{2022A&A...657A..48S} point out that even for field systems, the multiplicity fraction is less dependent on primary mass than it is commonly expected when limited range of mass ratios is considered.}

The pre-Gaia estimate is consistent with our measurements: 
$f=23\pm6$\% for $q>0.5$ \citep{2003MNRAS.342.1241P}. Studies based solely on Gaia data tend to underestimate the binary fraction, which we attribute partly to the predominant exclusion of binary stars during membership classification due to problematic astrometric solutions: 
 \begin{itemize}
     \item $f=8.6\pm1.2\%$ \citep{2023A&A...675A..89D}, $14\pm2\%$ \cite{2021AJ....162..264J}, $15\pm10$\% \citep{2023A&A...672A..29C} for $q>0.6$
 \item $f=8.7$\% \citep{2024arXiv241023527A}, $13$\% \citep{2024ApJ...971...71J} for $q> 0.5$
   \item $f=18.8\pm0.5$\% \citep{2023AJ....166..110P},  $22.9\pm2.6$\% \citep{2025MNRAS.536..471A} for $q>0.4$ 
   \end{itemize}

The obtained values increase when data on spectroscopic binarity is implemented:  
$f=20\pm3$\% for $q>0.5$ \citep{2020ApJ...903...93N} and $f=22.4\pm4.2$\% for $q>0.6$ \citep{2023ApJS..268...30L}. The latter study benefits from the inclusion of high-RUWE Gaia sources, which are often discarded.
 
Following the spectroscopic survey, \cite{2021ApJ...921..117T} obtained the binary fraction of $25\pm3$\% for periods within $10^3$ days or $42\pm4$\% for $P<10^4$ days, and at least 57\% for all binaries. If flat mass ratio distribution holds (figure 15 in the respective paper), the fraction for ${q<0.5}$ pairs should be a half of that, being close to 30\%.


\section{Conclusions}
\label{summary}

Following a resolved multiplicity survey of 409 Pleiades stars with speckle interferometric observations and Gaia, we draw the following conclusions:

\begin{itemize}
    \item The Gaia multipeak fraction is a strong predictor of subarcsecond multiplicity; all sources with ${\it ipd\_frac\_multi\_peak} >4\%$ are resolved as binary or multiple systems (Section \ref{resolved_types}, Figure \ref{fig:two-color}). 
    \item We report the discovery of 21 stellar components in 20 systems with speckle interferometric observations. One triple and 13 binary stars were revealed among previously unresolved stars, and 6 pairs were upgraded to triple systems (Section \ref{discoveries}).
    \item The companion of Merope (23 Tau) was reobserved 21.5 years after its discovery, following several unsuccessful attempts (Section \ref{Merope}). 
      \item \textcolor{black}{A deficit of wide pairs with $s>300$ au is confirmed  with estimated density three times lower than in the field (Section \ref{Wide}, Figure \ref{fig:results}). The dearth of $q>0.55$ systems in 270 -- 1350 au projected separation range may indicate disruption of primordial binaries following cluster dynamical evolution.}
    \item \textcolor{black}{For 10\% of Pleiades stars, a companion with a mass ratio $q>0.5$ is found within 27~--~1350~au. The extrapolated binary fraction across all distances for $q>0.5$ pairs reaches 25\% (Section \ref{fraction})}. 
  \end{itemize}

\section*{Data availability}
Contrast curve and autocorrelation function plots for all observations, along with the full versions of Tables \ref{tab:binarity_observations} and \ref{tab:detection_limits}, are available at \href{https://doi.org/10.5281/zenodo.14252721}{10.5281/zenodo.14252721}. Table data are also available through the \href{http://vizier.cds.unistra.fr/viz-bin/VizieR?-source=J/AJ/169/145}{VizieR} service.

\section*{Acknowledgments} 
Dedicated to the memory of our friend and colleague, Dmitry Kolesnikov (1990 June 22 -- 2024 December 28). \\\\
We owe the engineering and scientific staff who maintain the SAI MSU observatory, particularly the observers Ivan Gerasimov, Alexander Tarasenkov, Polina Budnikova, Dmitry Cheryasov, Vera Postnikova, and Ivan Shaposhnikov. The referee's report that was full of useful comments and suggestions helped us to improve the paper. We acknowledge the Russian Telescope Time Allocation Committee which provided a generous amount of time for our proposal. The authors honor the local communities around the Shatdzhatmaz plateau which hosts the Caucasian Mountain Observatory of Sternberg Astronomical Institute. This study was carried out using equipment funded by the Program of the Development of \mbox{M.V. Lomonosov Moscow State University}.


%

\software{TOPCAT \citep{2005ASPC..347...29T}, The SIMBAD astronomical database \citep{2000A&AS..143....9W}, VizieR catalogue access tool \citep{2000A&AS..143...23O}, Aladin Lite \citep{2014ASPC..485..277B, 2022ASPC..532....7B}, the CDS Cross-match Service \citep{2012ASPC..461..291B, 2020ASPC..522..125P},  SAO/NASA Astrophysics Data System}


\newpage
\setlength{\textheight}{664.75pt}

\bibliography{sample631}{}
\bibliographystyle{aasjournal}

\setlength{\textheight}{657.3189pt}






\appendix

\section{Controversial cluster membership }
\label{Membership}

The following objects were excluded from statistics as likely non-members, though we retain ambiguity since their positions in the color-magnitude diagram are compatible with cluster membership \citep{2024AJ....168..156C}.

\textbf{WDS 03532+2356}. $G_1=12.57$, $G_2=15.30^{\rm mag}$. The primary has $\delta \mu= 9.8$ mas yr$^{-1}$ (6.3 km s$^{-1}$) relative proper motion compared to the cluster's average, and an 8~km~s$^{-1}$ radial velocity offset according to Gaia DR3. The parallax, $\varpi=8.43\pm0.01$~mas, is far from the mean $\varpi^c\sim7.35$~mas.

\textbf{WDS 03478+2233}. $G=14.51^{\rm mag}$. A two-parameter solution in Gaia, which lacks proper motion. However, ${\delta \mu= 12.6}$~mas~yr$^{-1}$ (8~km~s$^{-1}$) according to the PPMXL catalog \citep{2010AJ....139.2440R} data. The radial velocity differs by 55 km s$^{-1}$ from the mean.

\textbf{WDS 03493+2404}. $G=14.59^{\rm mag}$. A two-parameter solution in Gaia, $\delta \mu=20.9$ mas yr$^{-1}$ (13~km~s$^{-1}$) according to the PPMXL. The radial velocity differs by 40 km s$^{-1}$ in Gaia DR3 but agrees within 1 km s$^{-1}$ in APOGEE \citep{2022ApJS..259...35A}.  

\section{Multiple systems}
\label{Multiples}

We identify one sextuple, one quadruple, and 14 triple systems in our sample. The definition of the mass ratio is not straightforward for these systems. As a rule, the mass of the primary component in an unresolved pair is considered. \textcolor{black}{For example, if a remote companion C orbits an unresolved AB pair, the mass ratio is defined as ${q=m_C/m_A}$.} Data on spectroscopic binarity are from \cite{2021ApJ...921..117T}. 
For double-lined systems (SB2), we search the isochrone grid for systems whose total flux agrees with the reported $G$ magnitude and $q$ derived from the spectroscopic orbital solution. For single-lined binaries (SB1), only the lower limit of the secondary component's mass is known, as it depends on orbital inclination $i$. The source's position in the color-magnitude diagram is used to place additional constraints.

\begin{table*}[t!]
    \centering
        \caption{Sextuple system WDS 03500+2351. Nomenclature follows The Updated Multiple Star Catalog \citep{2018ApJS..235....6T}.        Three sources are resolved in Gaia DR3. For the SB2 pair, $q$ is well-defined, with larger ambiguity for the SB1 \citep{2021ApJ...921..117T}. The speckle-resolved component Cb does not contribute to $G_3$ magnitude. Since source C is unresolved in $B_p$ and $R_p$ passbands, low flux from the $C_{a2}$ component is predicted. A uniform distribution in 0.4 -- 0.6 $M_\odot$ range for its mass is adopted.} 
    \begin{tabular}{|c|c|c|c|c|c|c|}
        \hline
Parameter& \multicolumn{6}{c|}{WDS 03500+2351 components}\\ \hline
 Source& \multicolumn{2}{c|}{A}&B& \multicolumn{3}{c|}{C}\\ \hline
 Gaia DR3&\multicolumn{2}{c|}{66507469798631808}&66507469798631936&\multicolumn{3}{c|}{66507469798632320} \\ \hline
 Separation $\rho$, $^{\prime\prime}$&\multicolumn{2}{c|}{0.00}&3.28&\multicolumn{3}{c|}{10.11}\\ \hline
 Observed $G$&\multicolumn{2}{c|}{$G_1=6.81^{\rm mag}$}&$G_2=10.01^{\rm mag}$&\multicolumn{3}{c|}{$G_3=10.23^{\rm mag}$}\\  \hline
RUWE &\multicolumn{2}{c|}{0.974}&1.116&\multicolumn{3}{c|}{11.596}\\ \hline
 Component&Aa&Ab&B&Ca1&Ca2&Cb     \\ \hline
 Type&\multicolumn{2}{c|}{SB2}&--&\multicolumn{2}{c|}{SB1, $m_{\rm C_{a2}}\sin i= 0.40\pm 0.01\ M_\odot$}& Speckle, $\rho=0.53^{\prime\prime}$\\ \hline
 Mass, $M_\odot$&$2.42 \pm 0.12$&$1.33\pm0.06$&$1.15\pm 0.04$&$1.10\pm0.03$  & $0.50 \pm 0.07$&$0.62\pm0.05$ \\ \hline
 Mass ratio& \multicolumn{2}{c|}{$m_{\rm Ab}/m_{\rm Aa}$}&$m_{\rm B}/m_{\rm Aa}$&$m_{\rm Ca1}/m_{\rm Aa}$&$m_{\rm Ca2}/m_{\rm Ca1}$&$m_{\rm Cb}/m_{\rm Ca1}$ \\  
$q$&\multicolumn{2}{c|}{$0.549\pm0.001$}&$0.48\pm0.03$&$0.45\pm0.03$&$0.45\pm0.07$&$0.57\pm0.05$
\\  
 \hline
    \end{tabular}
    \label{tab:hd23964}
\end{table*}

\textbf{WDS 03500+2351}. $G_1=6.81$, $G_2=10.01$, ${G_3=10.23^{\rm mag}}$. 
Probably sextuple system (Table~\ref{tab:hd23964}).

\textbf{WDS 03509+2358}. $G_1=6.93$, $G_2=15.96^{\rm mag}$, ${\rho=2.8^{\prime\prime}}$. For a pure binary, these magnitudes correspond to $q=0.15$, but the primary source's position in the color-magnitude diagram and its large RUWE  \textcolor{black}{(4.034)} suggest a close unresolved system. In case of a twin inner binary, the outer pair's $q$ increases to $0.19$. 

\textbf{WDS 03456+2420}. $G_1=7.25$, $G_2=15.45^{\rm mag}$, $\rho=5.3^{\prime\prime}$. A triple system, with the brighter source speckle-resolved at $\rho=0.26^{\prime\prime}$. The estimated masses are $m_{\rm Aa}=2.18 \pm 0.11$ and $m_{\rm Ab}=1.42 \pm 0.12\ M_\odot$ for inner pair, and $m_{\rm B}=0.43\pm0.02\ M_\odot$ for distant companion. 

\textbf{WDS 03474+2355}. $G_1=7.29$, ${G_2=9.74^{\rm mag}}$, $\rho=6.3^{\prime\prime}$. Quadruple system. The brighter source is an SB1 system. Companion with ${m_{\rm Ab} \sin i=0.12\ M_\odot}$ has negligible flux, and the primary mass is ${m_{\rm Aa}=2.16\pm0.11\ M_\odot}$. The fainter Gaia source is SB2 pair with $q=0.930\pm0.015$ and component masses ${m_{\rm Ba}=1.09 \pm 0.03}$ and $m_{\rm Bb}=1.01 \pm 0.03\ M_\odot$. 

\textbf{WDS 03487+2316}. $G_1=8.34$, $G_2=17.39^{\rm mag}$, $\rho=5.0^{\prime\prime}$. Triple system, the primary component is speckle-resolved with $\rho=0.85^{\prime\prime}$. The estimated masses are $m_{\rm Aa}=1.62 \pm 0.06$, $m_{\rm Ab}=0.59 \pm 0.04\ M_\odot$ for inner pair and $m_{\rm B}=0.20\pm0.02\ M_\odot$ for a remote companion.

\textbf{WDS 03475+2406} D. $G=8.62^{\rm mag}$. Triple-lined spectroscopic system, $q=0.956\pm0.005$ for unresolved  inner pair. The speckle-resolved component $D_{\rm b}$ at $\rho=0.15^{\prime\prime}$ likely does not contribute to the reported Gaia magnitude, given its multipeak excess and offset in the color-color diagram (Section \ref{resolved_types}). Component A in WDS is Alcyone, which is 2.2 arcmin apart. Estimated masses are ${m_{\rm Da1}=1.35\pm0.04}$, $m_{\rm Da2}= 1.25\pm0.04$, ${m_{\rm Db}= 1.16\pm0.07\ M_\odot}$ 

\textbf{WDS 03444+2408},  $G_1=8.97$, $G_2=14.21^{\rm mag}$, ${\rho=3.8^{\prime\prime}}$. 
Primary source is speckle-resolved with ${\rho=0.08^{\prime\prime}}$, masses in the inner pair are estimated as $M_{\rm Aa}=1.40\pm0.05$, $M_{\rm Ab}=0.71\pm0.06\ M_\odot$, the remote companion has $M_{\rm B}=0.56\pm0.01\ M_\odot$.

\textbf{WDS 03434+2314}. $G_1=10.36$, $G_2=15.69^{\rm mag}$, $\rho=3.6^{\prime\prime}$. Speckle-resolved companion at $\rho=0.11^{\prime\prime}$. Stellar masses for inner pair are $m_{\rm Aa}=1.07\pm0.03$ and $m_{\rm Ab}=0.61\pm0.04\ M_\odot$, $m_{\rm B}=0.40\pm 0.02\ M_\odot$ for outer component.

\textbf{WDS 03488+2416}. $G=10.71^{\rm mag}$. A triple-lined system, with the short-period ($P=49$ d) Ba~--~Bb pair  orbiting the brighter star A over $P \sim 37$~yr. Completely unresolved source in Gaia DR3, the wider pair is speckle-resolved with $\rho=0.1^{\prime\prime}$. Mass ratio ${(m_{Ba}+m_{Bb})/m_A=1.49\pm0.06}$ from spectroscopic solution. The derived masses are $m_{\rm A}=0.92\pm0.03$, $m_{\rm Ba}=0.82\pm0.06$, $m_{\rm Bb}=0.55\pm0.09\ M_\odot$. 

\textbf{WDS 03453+2517}. $G_1=11.01$, $G_2=17.27^{\rm mag}$, $\rho=3.9^{\prime\prime}$. Primary Gaia source is a SB1 system with low-mass $m_{\rm Ab} \sin i=0.295 \pm 0.001\ M_\odot$ companion. Small offset from the main sequence is probably noticeable in the color-magnitude diagram. \textcolor{black}{We allow a uniform distribution within the 0.3 -- 0.55 $M_\odot$ range for its mass, as a larger companion will impact the observed photometry significantly.} Stellar masses are ${m_{\rm Aa}=0.95\pm0.03}$, $m_{\rm Ab}=0.42\pm0.09$, $m_{\rm B}=0.21\pm0.02\ M_\odot$.

\textbf{WDS 03441+2402}. $G_1=11.17$, $G_2=11.86^{\rm mag}$, $\rho=5.7^{\prime\prime}$. The secondary component is speckle-resolved at $\rho=0.25^{\prime\prime}$. The masses are $m_{\rm A}=0.93 \pm 0.03$, ${m_{\rm Ba}=0.82\pm0.02}$, $m_{\rm Bb}=0.44\pm0.04\ M_\odot$. 

\textbf{WDS 03447+2553}. $G=11.44^{\rm mag}$. Previously unreported companions at $\rho =0.58$ and $0.66$ arcsec form a close pair with $\rho=0.13^{\prime\prime}$. The component's masses are ${m_{\rm A}=0.89\pm0.03}$, $m_{\rm B}=0.26\pm0.04$, ${m_{\rm C}=0.25\pm0.03\ M_\odot}$.

\textbf{WDS 03520+2440}. $G=11.73^{\rm mag}$. The inner pair, with $\rho=0.11^{\prime\prime}$ and masses $m_{\rm A}=0.84\pm0.02$, ${m_{\rm B}=0.80\pm0.04\ M_\odot}$, is accompanied by a remote companion with $m_{\rm C}=0.60\pm0.04\ M_\odot$ at $\rho=0.42^{\prime\prime}$.  

\textbf{WDS 03467+2456}, $G=13.44^{\rm mag}$. Discovered as a binary with $\rho=0.4^{\prime\prime}$ by \cite{2018AJ....155...51H}. We reveal the companion forms a twin pair with ${\rho=0.14^{\prime\prime}}$. Component masses are $m_{\rm A}=0.62\pm0.02$ and ${m_{\rm Ba}=m_{\rm Bb}=0.24\pm0.03\ M_\odot}$.

\textbf{WDS 03427+2412}, $G_1=13.85$, $G_2=16.35^{\rm mag}$, $\rho=1.7^{\prime\prime}$. Secondary Gaia source forms a close pair with $\rho=0.21^{\prime\prime}$ and $\Delta I_c=0.2^{\rm mag}$. The estimated mass of the primary is $m_{\rm A}=0.58\pm 0.01\ M_\odot$, and $m_{\rm Ba}=0.24\pm0.02$, $m_{\rm Bb}=0.21\pm0.02\ M_\odot$ for the companions.

\textbf{WDS 03459+2552}. $G_1=14.52$, $G_2=15.66^{\rm mag}$, $\rho=1.6^{\prime\prime}$. 
A simple binary according to our observations and Gaia, however
secondary source was resolved by \cite{1997A&A...323..139B} with magnitude difference $\Delta J= 0.13\pm0.01^{\rm mag}$, implying $q=0.92\pm0.01$. Primary mass is $m_A=0.53\pm0.01\ M_\odot$, secondary companions have $m_B=0.32\pm0.02$ and $m_C=0.29\pm0.02\ M_\odot$.

\onecolumngrid
\newpage
\section{Binarity Observations and Detection Limits}
\label{log}
\startlongtable
\begin{deluxetable*}{ccccccccccc}
\vspace{-0.5cm}
\tabletypesize{\scriptsize}
\tablecaption{
Observational log for speckle-resolved Pleiades stars.
\label{tab:binarity_observations}}
\tablewidth{0pt}
\tablehead{ \colhead{Gaia DR3} & \colhead{WDS} & \colhead{G}& \colhead{Date}& \colhead{Band} & \colhead{$\beta$} & \colhead{Pair} & \colhead{$\rho$}& \colhead{$\theta$} & \colhead{$\epsilon$}  & Notes \\
             \colhead{Source} &  & \colhead{mag}& \colhead{yr}&  & \colhead{${}^{\prime\prime}$} & & \colhead{${}^{\prime\prime}$}& \colhead{deg} & $\Delta \mathrm{mag}$ & 
}
\colnumbers
\startdata
65205373152172032&03463+2357&4.173&2023.846&$\hat{y}$&1.12&\romannumeral 1&0.2676$\pm$0.0025&53.97$\pm$0.54&5.63$\pm$0.12&\\
65205373152172032&03463+2357&4.173&2023.846&$\hat{z}$&1.03&\romannumeral 1&0.2784$\pm$0.0032&54.20$\pm$0.65&4.85$\pm$0.12&\footnotesize{q}\\
65205373152172032&03463+2357&4.173&2024.251&$\hat{y}$&1.04&\romannumeral 1&0.2889$\pm$0.0099&54.01$\pm$1.95&5.99$\pm$0.21&\footnotesize{R}\\
65205373152172032&03463+2357&4.173&2024.251&$\hat{z}$&0.94&\romannumeral 1&0.2694$\pm$0.0276&58.01$\pm$5.86&5.63$\pm$0.42&\footnotesize{R}\\
65282716922610944&03456+2420&7.249&2023.846&$\hat{y}$&1.01&\romannumeral 1&0.2602$\pm$0.0007&179.28$\pm$0.14&1.69$\pm$0.01&\\
65282716922610944&03456+2420&7.249&2023.846&$\hat{z}$&0.92&\romannumeral 1&0.2618$\pm$0.0007&179.13$\pm$0.14&1.24$\pm$0.01&\footnotesize{q}\\
65282716922610944&03456+2420&7.249&2024.251&$\hat{y}$&1.06&\romannumeral 1&0.2630$\pm$0.0007&179.28$\pm$0.14&1.74$\pm$0.01&\\
65282716922610944&03456+2420&7.249&2024.251&$\hat{z}$&1.11&\romannumeral 1&0.2611$\pm$0.0007&179.05$\pm$0.15&1.22$\pm$0.01&\\
66724451545088128&03482+2419&8.214&2024.163&$\hat{y}$&1.0&\romannumeral 1&0.0724$\pm$0.0002&344.02$\pm$0.19&1.65$\pm$0.01&\footnotesize{q}\\
66724451545088128&03482+2419&8.214&2024.163&$\hat{z}$&0.85&\romannumeral 1&0.0734$\pm$0.0004&342.18$\pm$0.34&1.32$\pm$0.01&\\
66724451545088128&03482+2419&8.214&2024.240&$R_c$&1.19&\romannumeral 1&0.0652$\pm$0.0002&340.55$\pm$0.18&1.21$\pm$0.01&\footnotesize{R},\footnotesize{P}\\
66724451545088128&03482+2419&8.214&2024.240&$I_c$&1.26&\romannumeral 1&0.0654$\pm$0.0003&341.03$\pm$0.24&1.00$\pm$0.01&\footnotesize{R},\footnotesize{P}\\
66724451545088128&03482+2419&8.214&2024.251&$\hat{y}$&1.0&\romannumeral 1&0.0740$\pm$0.0005&342.75$\pm$0.38&1.62$\pm$0.04&\footnotesize{R}\\
\enddata 
\tablecomments{Columns: (1) Gaia DR3 designation of primary star; (2)  WDS designation; (3) $G$ magnitude; (4) Observation date; (5) Passband; (6) Seeing; (7) Pair number within multiple system; (8)  Angular separation; (9) Position angle; (10) Contrast; (11) Notes. \\ 
P: 180 degree ambiguity in position angle. \\ 
R: No reference star used; contrast estimation less accurate. \\ 
U: Unreliable contrast due to anisoplanatism. \\ 
q: Observation used for $q$ calculation. \\ 
W: WDS designation newly introduced. \\ 
(This table is available in its entirety in machine-readable form at \href{https://doi.org/10.5281/zenodo.14252721}{10.5281/zenodo.14252721} and through the \href{http://vizier.cds.unistra.fr/viz-bin/VizieR-3?-source=J/AJ/169/145/table5}{VizieR} service)}
\end{deluxetable*}

\onecolumngrid

\startlongtable
\vspace{-0.5cm}
\begin{deluxetable*}{ccccccccccc}
\tabletypesize{\footnotesize}
\tablecaption{
Detection limits for Pleiades sample stars.
\label{tab:detection_limits}}
\tablewidth{0pt}
\tablehead{ \colhead{Gaia DR3} & \colhead{G}& \colhead{Date}& \colhead{Band} & \colhead{$\beta$} & \colhead{$\epsilon_{lim_{0.1}}$} & \colhead{$\epsilon_{lim_{0.2}}$}& \colhead{$\epsilon_{lim_{0.4}}$} & \colhead{$\epsilon_{lim_{0.6}}$} & \colhead{$\epsilon_{lim_{1.0}}$} & Notes \\
             \colhead{Source} &  \colhead{mag}& \colhead{yr}&         & \colhead{${}^{\prime\prime}$} & \colhead{$\Delta \mathrm{mag}$} & \colhead{$\Delta \mathrm{mag}$} & \colhead{$\Delta \mathrm{mag}$} & \colhead{$\Delta \mathrm{mag}$} & \colhead{$\Delta \mathrm{mag}$} & 
}
\colnumbers
\startdata
66714384142368256&2.896&2023.920&$I_c$&1.07&2.40&3.98&5.78&5.77&6.09&\\
66714384142368256&2.896&2023.920&$\hat{y}$&1.36&2.76&5.44&6.72&6.38&6.90&\\
66714384142368256&2.896&2023.920&$\hat{z}$&1.20&1.26&3.71&5.52&5.60&6.02&\\
66526127137440128&3.616&2023.920&$\hat{y}$&1.39&2.40&5.25&6.34&6.36&6.96&\\
66526127137440128&3.616&2023.920&$\hat{z}$&1.13&1.52&3.56&5.19&5.69&6.34&\\
65271996684817280&3.698&2023.846&$\hat{y}$&0.91&3.52&5.68&6.98&6.99&8.28&\\
65271996684817280&3.698&2023.846&$\hat{z}$&0.97&3.29&4.76&6.00&6.27&7.76&\\
65283232316451328&3.863&2023.846&$\hat{z}$&0.84&1.87&3.98&5.48&5.79&7.52&\\
65283232316451328&3.863&2023.846&$\hat{y}$&0.97&3.00&5.64&6.25&6.31&8.23&\\
65205373152172032&4.173&2024.251&$\hat{z}$&0.94&1.91&4.13&5.50&6.04&7.07&b\\
65205373152172032&4.173&2024.251&$\hat{y}$&1.04&3.41&5.75&6.59&6.85&8.11&b\\
\enddata 
\tablecomments{Columns: (1) Gaia DR3 source designation; (2) $G$ magnitude; (3) Observation date; (4) Passband; (5) Seeing; (6-10) Detection limits at 0.1, 0.2, 0.4, 0.6, and 1.0 arcsec; (11) Notes. \\ 
b: System is successfully resolved, see binarity parameters in Table~\ref{tab:binarity_observations}. \\ 
d: Detection limits are affected (degraded) by the resolved component. \\ 
(This table is available in its entirety in machine-readable form at \href{https://doi.org/10.5281/zenodo.14252721}{10.5281/zenodo.14252721} and via the \href{http://vizier.cds.unistra.fr/viz-bin/VizieR-3?-source=J/AJ/169/145/table6}{VizieR} service)}
\end{deluxetable*}



\end{document}